\documentclass[article]{jss}

\usepackage{bm}
\usepackage{amsmath,amssymb}
\usepackage{latexsym}
\usepackage{graphicx}
\usepackage[T1]{fontenc}
\usepackage{bm}
\usepackage{extarrows}
\usepackage{fancyvrb}
\usepackage{tikz}
\usepackage{multicol}
\usepackage{multirow}
\usepackage{listings}
\usepackage{indentfirst}
\usepackage{color}
\usepackage{setspace}
\usepackage{textcomp}
\usepackage{url}
\usepackage{hyperref}
\usepackage[final]{pdfpages}
\usepackage{pmboxdraw}

\definecolor{mygreen}{rgb}{0,0.6,0}
\definecolor{mygray}{rgb}{0.5,0.5,0.5}
\definecolor{mymauve}{rgb}{0.58,0,0.82}

\author{Peiran Liu\\University of Washington \And Hana {\v{S}}ev{\v{c}}{\'\i}kov{\'a} \\University of Washington \And
        Adrian E. Raftery\\University of Washington}
\title{Probabilistic Estimation and Projection of the Annual Total Fertility Rate Accounting for Past Uncertainty:
A Major Update of the \pkg{bayesTFR} \proglang{R} Package.
}
\Plainauthor{Peiran Liu, Hana {\v{S}}ev{\v{c}}{\'\i}kov{\'a}, Adrian E. Raftery} %% comma-separated
\Plaintitle{A Major Update of the R Package bayesTFR: Probabilistic Estimation and Projections of the Annual Total Fertility Rate Accounting for Past Uncertainty} %% without formatting
\Shorttitle{\pkg{bayesTFR} Update: Annual TFR with Uncertainty} %% a short title (if necessary)

\Abstract{
  The \pkg{bayesTFR} package for \proglang{R} provides a set of functions to produce probabilistic projections of the total fertility rates (TFR) for all countries, and is widely used, including as part of the basis for the UN's official population projections for all countries. \cite{liu2018accounting} extended the theoretical model  by adding a layer that accounts for the past TFR estimation uncertainty. A major update of \pkg{bayesTFR} implements the new extension.  Moreover, a new feature of producing  annual TFR estimation and projections extends the existing functionality of estimating and projecting for five-year time periods. An additional autoregressive component has been developed in order to account for the larger autocorrelation in the annual version of the model. This article summarizes the updated model, describes the basic steps to generate probabilistic estimation and projections under different settings, compares performance, and provides instructions on how to summarize, visualize and diagnose the model results.
}
\Plainkeywords{bayesTFR, autoregressive model,  Bayesian hierarchical model, Markov chain Monte Carlo, R, United Nations, World Population Prospects, annual projections, past TFR uncertainty} %% without formatting
\Address{
  Peiran Liu  \\
  Department of Statistics\\
  University of Washington\\
  Seattle, WA, United States of America\\
  E-mail: \email{prliu@uw.edu}\\
  
  Hana {\v{S}}ev{\v{c}}{\'\i}kov{\'a}  \\
  Center for Statistics and the Social Sciences\\
  University of Washington\\
  Box 354322 \\
  Seattle, WA 98195-4322, United States of America\\
  E-mail: \email{hanas@uw.edu}\\
  
  Adrian E. Raftery  \\
  Department of Statistics and Sociology\\
  University of Washington\\
  Box 354320 \\
  Seattle, WA 98195-4320, United States of America\\
  E-mail: \email{raftery@uw.edu}\\
  URL: \url{https://www.stat.washington.edu/raftery/}
}

\begin{document}

%% include your article here, just as usual
%% Note that you should use the \pkg{}, \proglang{} and \code{} commands.

\section[Intro]{Introduction}
%% Note: If there is markup in \(sub)section, then it has to be escape as above.
In 2015 for the first time, the United Nations (UN) adopted the Bayesian method described by \cite{alkema2011probabilistic} for their official population projections for all countries, the World Population Prospects (WPP)~2015 \citep{UN2015}. This method is probabilistic and based on a principled statistical footing, replacing the previous deterministic method. One of the major components is the projection of the total fertility rate (TFR) which is implemented in the \pkg{bayesTFR} \proglang{R} package~\citep{vsevvcikova2011bayestfr}. This package is widely used in research on fertility rates and population projections (\cite{abel2016meeting, gerland2017patterns, vsevvcikova2016bayespop, vsevvcikova2018probabilistic}).

While the projection of TFR is probabilistic, the method does not take uncertainty about the past into account. \cite{liu2018accounting} addressed this issue by developing a Bayesian model that takes past TFR observations from the World Fertility Data database \citep{UN2019WFD} as raw data, and combines the uncertainty from the data with the uncertainty from the model. Out-of-sample validation showed improved performance of the overall projection model, while providing users with information about the uncertainty of estimates of past fertility rates. A major overhaul of \pkg{bayesTFR} was required to incorporate the \citet{liu2018accounting} methodology into the package.

The original framework implemented in \pkg{bayesTFR} was designed to work with five-year estimates and produced projections on a five-year time interval basis. This has the disadvantage of missing TFR fluctuations and pattern changes within the five-year periods. There is a growing interest by the UN to publish population estimates and projections on an annual basis, and in response we have extended \pkg{bayesTFR} to work with annual data. The update revealed that an additional autoregressive component is needed to account for the larger autocorrelation and thus, to model the uncertainty in the fertility transition well. 

The new version of the package, version 7.1-1~\citep{bayesTFRpackage}, now produces uncertainty information about the past which is propagated into the projections and is able to estimate and project on an annual basis. This article describes the methodological changes, and also provides instructions on how to generate probabilistic estimations and projections under different settings. These include with and without accounting for past TFR estimation, with annual or five-year data, and with and without the autoregressive component in the fertility transition phase of the model. Other updates to the package are also introduced and elaborated. %The new version is currently available on the GitHub repository \code{PPgp/bayesTFR}, branch \code{dev}.  
%The package is available from the Comprehensive \proglang{R} Archive Network at \url{http://CRAN.R-project.org/package=bayesTFR} with version 7.0-0.

The article is organized as follows. Section~\ref{sec2} summarizes the theoretical models developed by \cite{alkema2011probabilistic, raftery2014bayesian, liu2018accounting}, and the autoregressive model in the fertility transition phase. Section~\ref{sec3} describes how to use the package, using a step-by-step approach with different model settings. Section~\ref{experiments} presents experiments on the performance of the models and the selection of the various settings. The article concludes with a discussion in Section~\ref{discussion}.

\section[Method]{Annual TFR model with uncertainty about the past}\label{sec2}
Here, we first summarize the original TFR model  developed for five-year time periods \citep{alkema2011probabilistic}. We then review the new methodology for probabilistic estimation and projection of TFR for all countries of the world accounting for uncertainty about the past, as proposed by \cite{liu2018accounting}. Finally, we describe the changes in the methodology to work for annual estimation and projections.

TFR can be defined as the number of children a woman would have if she were subject to the prevailing fertility rates at all ages from a single given year, and survived throughout her childbearing years. 
\cite{alkema2011probabilistic} defined a three-phase model for the evolution
of TFR over time in a country:
\begin{enumerate}
\item Phase I: pre-transition phase with fluctuations at high fertility level.
\item Phase II: transition from high to low fertility, where decrements are modeled by a random walk with drift given by a double logistic function.
\item Phase III: post-transition phase where fertility fluctuates around the replacement level (a level close to 2.1), modeled by an autoregressive AR(1) process.
\end{enumerate}

We will use the same notation as \cite{vsevvcikova2011bayestfr}. Specifically, $f_{c,t}$ denotes the TFR in country $c$ and time period $t$, $\tau_c$ denotes the start period of Phase II for country $c$, $\lambda_c$ is the start period of Phase III for country $c$, while $g(\bm{\theta}_c, f_{c,t})$ and $\bm{\theta}_c$ denote the parametric decline function and the corresponding country-specific parameters, respectively.

\subsection{Existing model with five-year estimates}\label{existing}
The pre-transition phase (Phase I) is not modeled, as all countries have already entered Phase II. Thus, for the purpose of projecting into the future it is not needed.

The fertility transition phase (Phase II) is modeled by a random walk with drift. This is specified by
\begin{align}\label{model1}
& f_{c,t+1} = f_{c,t} - d_{c,t} \quad \text{for} \quad \tau_c \leq t < \lambda_c\,.
\end{align}

The decrement $d_{c,t}$ in (\ref{model1}) is modeled as the sum of a function of the level of the TFR and the noise, as follows:
\begin{align}\label{model:old}
& d_{c,t} = d(\bm{\theta}_c, \lambda_c, \tau_c, f_{c,t}) = g(\bm{\theta}_c, f_{c,t}) + \varepsilon_{c,t}
\end{align}
where $g(\bm{\theta}_c, f_{c,t})$ are the double logistic decrements, which are determined by the country-specific parameter vector $\bm{\theta}_c = (\Delta_{c1},\Delta_{c2},\Delta_{c3},\Delta_{c4}, d_c)$ and given by
\begin{align}\label{dlcurve}
\frac{-d_c}{1+\exp\left(-\frac{2\ln(p_1)}{\Delta_{c1}}(f_{c,t}-\sum_{i=1}^4\Delta_{ci} + 0.5\Delta_{c1})\right)} + \frac{d_c}{1+\exp\left(-\frac{2\ln(p_2)}{\Delta_{c3}}(f_{c,t}-\Delta_{c4} - 0.5\Delta_{c3})\right)} .
\end{align}

The random distortions $\varepsilon_{c,t}$ in each period have normal distributions as follows:
\begin{align}\label{epsilon-PII}
\varepsilon_{c,t} \sim \left\{
\begin{aligned}
& N(m_t, s_t^2), &&\quad \text{for }t=\tau_c\,, \\
& N(0, \sigma(f_{c,t})^2) &&\quad \text{otherwise}\,.
\end{aligned}\right.
\end{align}
The quantity $ \sigma(f_{c,t})$ is the standard deviation of the distortions during the later periods with
\begin{align}\label{sigmaft}
& \sigma(f_{c,t}) = c_{1975}(t)\left(\sigma_0 + (f_{c,t}-S)(-aI_{[S,\infty)}(f_{c,t}) + bI_{[0,S]}(f_{c,t}))\right)\, .
\end{align}

The constant $c_{1975}(t)$ is added to model the higher error
variance of the distortions before 1975. For further details about the model and its priors, see \cite{vsevvcikova2011bayestfr}. For the purpose of this article, we only point to the definition of two parameters, namely the country-specific maximum decrement $d_c$, and the hyperparameter for the maximum standard deviation of the distortions $\sigma_0$. The $d_c$ parameter is defined as 
\begin{align}\label{dcstar}
& d_c^\ast = \log\left(\frac{d_c-0.25}{2.5-d_c}\right)\,,\\
& d_c^\ast \sim N(\chi, \psi^2)\,.\nonumber
\end{align}
The prior distribution of $\sigma_0$ is $\sigma_0 \sim U[0.01, 0.6]$.

The TFR in the post-transition phase (Phase III) is modeled by a first order autoregressive time series model \citep{raftery2014bayesian} as
\begin{align}\label{modelphase3}
f_{c,t+1} \sim N(\mu_c + \rho_c(f_{c,t}-\mu_c), \sigma^2) \text{ for }t\geq \lambda_c\,,
\end{align}
where $\mu_c$ is the country-specific long-term mean fertility rate, and $\rho_c$ is the autoregressive parameter with $\rho_c \in (0,1)$. In \pkg{bayesTFR} these parameters can be estimated  via the Markov chain Monte Carlo (MCMC) method. Alternatively, country-independent values can be pre-defined or estimated by maximum likelihood.

The start period of Phase II, $\tau_c$, is defined as
\begin{align}\label{tauc}
& \tau_c = \left\{\begin{aligned}
& \max\{t: (M_c-L_{c,t}) < 0.5\}, &\quad \text{if }L_{c,t} > 5.5;\\
& \,\text{first estimation year},&\text{otherwise,}
\end{aligned} \right.
\end{align}
where $M_c$ is the maximum observed TFR outcome in country $c$, and $L_{c,t}$ denote local maxima.

The start period of Phase III for country $c$, $\lambda_c$, is defined as the period where two consecutive increases of TFR below 2 have been observed. More formally,
\begin{align}\label{lambdac}
\lambda_c = \min \{ t: f_{c,t} > f_{c,t-1}, f_{c,t+1} > f_{c,t} \mbox{ and } f_{c,p} < 2  \text{ for } p = t-1, t, t+1 \}\,.
\end{align}

%\subsection{Model extensions and updates}\label{extension}
\subsection{Probabilistic TFR estimation with uncertainty}
\label{sec:uncertainty}
The method described above uses observed TFR values as input to estimate the model parameters. In the UN context, these input values are taken from the latest revision of the WPP. 
Such TFR data are in fact estimates of the observed values, often derived from multiple data sources and involve varying amounts of uncertainty. The TFR model from the previous section however, treats these estimates as true values.   

\cite{liu2018accounting} developed a method that assesses the uncertainty around past estimates of the observed TFR values and propagates it into the projections. The medians of the resulting posterior distributions can be used as point estimates of the past TFR, reducing the need for manual analysis and assessments by individual UN analysts. In addition, TFR projections resulting from the method of \cite{liu2018accounting} show better out-of-sample validation, especially better coverage of the prediction intervals, than the existing method.

The method uses the World Fertility Data database \citep{UN2019WFD} for past raw TFR estimates from surveys, reports and vital registrations for most regions in the world. We denote these data points by $y_{c,t,s}$, i.e., the raw TFR estimate for country $c$, time $t$ and source $s$. The source $s$ may refer to a census, a survey, vital registration statistics or other sources.  For each observation, there are features $\bm{x}_{c,s}$ that describe the sources, estimating methods, recall lags and other aspects of data collection and estimation, often measures of the quality of the data.  The observed $y_{c,t,s}$ are modeled as:
\begin{align}
& y_{c,t,s}|f_{c,t} \sim N(f_{c,t} + \delta_{c,s}, \rho_{c,s}^2) \label{model:unc} , \\
& E[\delta_{c,s}] = \bm{x}_{c,s}\bm{\beta} , \nonumber\\
& E[\rho_{c,s}] = \bm{x}_{c,s}\bm{\gamma} . \nonumber
\end{align}

The $\delta_{c,s}$ and $\rho_{c,s}$ are country-specific parameters which are estimated by maximum likelihood. In \cite{liu2018accounting}, the features used are the sources of the data and the corresponding estimation methods, but the model allows for any user-specified features. This part is combined with the existing Bayesian hierarchical model implemented in \pkg{bayesTFR}. Here, the past TFRs are considered as unknown, and are part of the parameters to estimate. The complete model is described in the Appendix.

If we are using TFR for five-year intervals, as for example in the \code{tfr} dataset  in the \pkg{wpp2019} package~\citep{wpp2019package}, the true TFR at any time stamp is considered to be the linear interpolation of two consecutive five-year TFRs, namely
\begin{align*}
& f_{c,t} = \frac{1}{5}[(t_{\ell+5}-t) f_{c,t_{\ell}} + (t - t_\ell)f_{c,t_{\ell+5}}]\quad \text{for any }t \in [t_{\ell}, t_{\ell+5}]\,.
\end{align*}

If we are estimating from annual TFR, for each observation of the raw data, we take the floor of $t$. For example, if an observation in the raw data is recorded at $1975.5$, we consider this observation as an estimate of the calendar year 1975.

Since we are now also modeling the past, not just the future as in the extant method, we need to model the pre-transition phase (Phase I), which is not necessary for projecting the future. The TFR in this phase will be modeled by a random walk, specified by
\begin{align*}
& f_{c,t+1} = f_{c,t} + \varepsilon_{c,t}\quad \text{for }t < \tau_c ,
\end{align*}
where the distributions of the random distortions in each period are given by
\begin{align*}
\varepsilon_{c,t} \sim N(0, s_t^2)\,.
\end{align*}
Here, we simplify the model by setting the variance to be the same as the variance in the first period of the TFR decrements. This is a reasonable assumption because the starting period of Phase~II is linked to Phase~I, and the expected decline of TFR at the starting period of Phase~II is small. Thus, the distortions of TFR share similar behavior.  

The estimation of all country-specific parameters and hyperparameters conditional on the TFRs, other than the TFRs themselves, in the Phase II model remains the same as described by \cite{vsevvcikova2011bayestfr}. To estimate past TFR, the model for Phase III is estimated together with the Phase II model via an MCMC algorithm \citep{gelfand1990sampling}. This algorithm is a combination of Gibbs sampling, Metropolis-Hastings (for $\Delta_{ci}$ in (\ref{dlcurve}) and TFR), and slice sampling steps \citep{neal2003slice}. 

The estimation yields a set of TFR trajectories about the past. To project into the future, we apply the existing projection method as described in \cite{vsevvcikova2011bayestfr} starting with the last estimation period of each trajectory. This is in contrast with the previous version, where the projection for each country starts from a single data point, namely the last observed TFR.

\subsection{Annual version of bayesTFR}
\label{sec:annual}
The original model described in Section~\ref{existing} was designed to work with five-year data. Several modifications needed to be made in order to estimate and project annual data well.

Most importantly, we found that unlike in the five-year version, the residuals of the Phase II model are highly autocorrelated when using annual data. We found that the lag 1 autocorrelation coefficients are about 0.7 for residuals of Phase II model for some major countries. Thus, we modified the Phase II model defined in Equations (\ref{model1})-(\ref{model:old}) by adding an additional first-order autoregressive component.  The random walk with drift model is then specified as
\begin{align}\label{model3}
& d_{c,t+1} - g(\bm{\theta}_c, f_{c,t+1}) = \phi(d_{c,t} - g(\bm{\theta}_c, f_{c,t})) + \varepsilon_{c,t} .
\end{align}

%The other important part is that with the five-year period TFR, the residuals are not significantly autocorrelated. However, if we assume that the residuals are independent with the following assumption,
%\begin{align*}
%& d_{c,t} = d(\bm{\theta}_c, \lambda_c, \tau_c, f_{c,t}) = g(\bm{\theta}_c, f_{c,t}) + \varepsilon_{c,t}
%\end{align*}
%The estimated residuals are significantly autocorrelated with correlation around 0.7 for major countries. This implies that if we sould build the fertility transition model as an order one autoregressive model with drift, which could be specified as:
%\begin{align}\label{model3}
%& d_{c,t+1} - g(\bm{\theta}_c, f_{c,t+1}) = \phi(d_{c,t} - g(\bm{\theta}_c, f_{c,t})) + \varepsilon_{c,t}
%\end{align}

The prior distribution of $\phi$ is set to be $\text{Uniform}(0,1)$ and is not country-specific. For the random distortions $\varepsilon_{c,t}$, the distribution is considered to be the same as in Equations~(\ref{epsilon-PII})-(\ref{sigmaft}).
%\begin{align*}
%& \varepsilon_{c,t} \sim N(0, \sigma(f_{c,t})^2), \quad\text{for }t>\tau_c \\
%& \sigma(f_{c,t}) = c_{1975}(t)\left(\sigma_0 + (f_{c,t}-S)(-aI_{[S,\infty)}(f_{c,t}) + bI_{[0,S]}(f_{c,t}))\right)
%\end{align*}

The same prior distributions as in the five-year version is used for most parameters. One exception is  $\sigma_0$ where the lower bound was decreased by a factor of the square root of five, i.e., $\sigma_0 \sim U[0.0045, 0.6]$. The upper bound was kept the same to allow for the possibility of additional correlation.

The definition of the maximum decrement defined in Equation~(\ref{dcstar}) was changed to be one-fifth of that for the five-year model:
\begin{align*}
& d_c^\ast = \log\left(\frac{d_c-0.05}{0.5-d_c}\right)\,.
\end{align*}

No changes have been made to the model of the post-transition phase of TFR, Phase III. It is modeled by a first-order autoregressive time series model as defined in Equation (\ref{modelphase3}).

The rule for determining the start period of Phase II, $\tau_c$, as defined in Equation~(\ref{tauc}), was unchanged. However, since the local maxima are calculated using annual TFR data, the results can differ from those obtained from a five-year dataset.

To determine the start periods of Phase III, $\lambda_c$, as defined in Equation~(\ref{lambdac}), we first obtain five-year averages of TFR and apply the same rule as in the five-year version, namely that Phase III starts when two consecutive increases of TFR below 2 are observed.

\subsection{Changes in TFR Projections}\label{sec:projections}
There are three main differences in the TFR projections between the new implementation and the one described by \cite{vsevvcikova2011bayestfr}.

The first difference (which we alluded to at the end of Section~\ref{sec:uncertainty}), relates to the fact that by accounting for the past TFR uncertainty (Equation~\ref{model:unc}), instead of observed point estimates, the model results in a set of TFR trajectories about the past which changes the starting values of the forecast. To project $f_{c,T+1}$ where $T$ is the last period of the estimation, the $i$-th sample from the MCMC output is given by $f_{c,T+1}^{(i)} = f_{c,T}^{(i)} - d_{c,T}^{(i)} + \varepsilon_{c,T}^{(i)}$. Thus, the uncertainty in the first forecast period will be wider than if we use a model without accounting for past uncertainty, in which case $f_{c,T}^{(i)} = f_{c,T}$ is the same for all trajectories.

The second difference relates to the annual model described in Section~\ref{sec:annual}. When the additional autocorrelation of Phase II is taken into account (Equation~\ref{model3}), the past noise is carried over to the next time period.  Specifically, to project $f_{c,t+1}$ for a country $c$ that is in Phase II at time $t$, the $i$-th sample is given by $f_{c,t+1}^{(i)} = f_{c,t}^{(i)} - d_{c,t}^{(i)} + \varepsilon_{c,t}^{(i)}$, where $d_{c,t}^{(i)} = g(f_{c,t}^{(i)}, \bm{\theta}_{c}^{(i)})$, and $\varepsilon_{c,t}^{(i)}$ is drawn from $N(\phi^{(i)}\varepsilon_{c,t-1}^{(i)}, \sigma^{(i)}(f_{c,t}^{(i)}))$. For the first forecast, i.e., at the time period $T+1$, the distortion of the last estimation period $T$ is calculated before starting the projections.

Finally, the last difference regards the updated Phase III model as described in \cite{raftery2014bayesian}, where country-specific long-term means $\mu_c$ and autocorrelation coefficients $\rho_c$  were incorporated into the model (Equation~\ref{modelphase3}) and estimated by MCMC. However, this change has been available in \pkg{bayesTFR} since version 3.0-0 was published in 2013.  Using this approach, to project $f_{c,t+1}$ for a country $c$ that is in Phase III at time $t$, the $i$-th MCMC sample is drawn from a Normal distribution $N\left(\mu_c^{(i)} + \rho_{c}^{(i)}(f_{c,t}^{(i)} - \mu_{c}^{(i)}), \sigma^{(i),2}\right)$.

\section{Overview of the Package Updates}\label{sec:update-overview}
The updated package \pkg{bayesTFR} retains all the functionalities of the previous version, which implements the model of \cite{alkema2012estimating}. Its new features  allow the user to conduct estimation of past TFR while accounting for uncertainty as described in \cite{liu2018accounting}, as well as to forecast TFR for both five-year and one-year time intervals, as requested by the UN. 

These new functionalities are implemented in the form of either new functions or additional arguments to existing functions. For convenience, especially for users who are familiar with the previous version of the package, this section summarizes the various changes. For users who are new to the package we recommend skipping to Section~\ref{sec3} where the usage is explained in more detail.
 
The following are established \pkg{bayesTFR} functions that were updated.
\begin{itemize}
	\item \code{run.tfr.mcmc}. This is the core function for MCMC estimation of fertility transition model parameters. The following optional arguments were added:
	\begin{enumerate}
		\item \code{annual}. Logical argument determining whether the model is trained based on annual TFR data (\code{TRUE}) or on the five-year data (\code{FALSE}). The default is \code{FALSE}.
		\item \code{ar.phase2}. Logical argument. If \code{TRUE}, model (\ref{model3}) will be used in the estimation, and the parameter $\phi$ will be estimated through the MCMC process. This is relevant only if \code{annual = TRUE}.
		\item \code{uncertainty}. Logical argument determining whether the model described in \cite{liu2018accounting} is estimated (\code{TRUE}) or if the default behavior of treating observed data as true values is used (\code{FALSE}). If \code{TRUE}, the past TFR values for all countries and time periods are estimated as additional parameters. Furthermore, Phase III of the TFR transition model is estimated simultaneously and thus, there is no need to call \code{run.tfr3.mcmc} separately. 
		\item \code{my.tfr.raw.file, covariates, cont_covariates, iso.unbiased}. These are arguments relevant to estimating past TFR. They allow the user to pass a file with raw TFR estimates, to set categorical and continuous covariates for estimating bias and measurement error variance of raw data, as well as to determine for which countries the vital registration TFR estimates are considered accurate. The arguments are considered only if \code{uncertainty = TRUE}.
	\end{enumerate} 

\item \code{tfr.predict}. This is the core function for TFR prediction. There was one optional argument added:
	\begin{itemize}
		\item \code{uncertainty}. Logical argument. If \code{TRUE} and the corresponding estimation was produced via \code{run.tfr.mcmc(..., uncertainty = TRUE)}, then each prediction trajectory starts from a trajectory representing the past. Otherwise all prediction trajectories start from the same point, namely the last observed TFR.
	\end{itemize}

\item \code{run.tfr.mcmc.extra}. Originally, this function has been implemented in order to estimate the TFR transition model for very small countries or countries with unusual historical patterns. These countries were excluded from \code{run.tfr.mcmc} in order not to bias the world parameters, and were estimated separately via this function. In this update, the function has been extended to recompute past TFR estimates on a country-specific basis. This allows users to analyze the impact of changes on the raw TFR of individual countries without needing to run a new simulation for the whole world. Added arguments to this function have the same meaning as for  \code{run.tfr.mcmc}:
	\begin{itemize}
		\item \code{uncertainty}, \code{my.tfr.raw.file}, \code{covariates} and \code{cont_covariates}, \code{iso.unbiased}
	\end{itemize}
\item \code{tfr.trajectories.table}, \code{tfr.trajectories.plot}. These functions give projection quantiles in tabular and graphical formats, respectively. They have been extended to include uncertainty information about the past, if such information exists.
\end{itemize}

The following are new functions added to the package. They are described in Section~\ref{sec:analresults} in more detail.
\begin{itemize}
\item \code{get.tfr.estimation}. Allows exploring the estimated trajectories as well as any quantiles of the past TFR estimates.
\item \code{tfr.estimation.plot}. Function for plotting estimates of past TFR for individual countries.
\item \code{tfr.bias.sd}. Allows exploring the bias and standard deviation estimated for the raw TFR estimates.
\end{itemize}

\section{Using the Updated bayesTFR}\label{sec3}
Previous versions of the \pkg{bayesTFR} package which implemented the model described in Section \ref{existing} have been used by UN analysts and others to train TFR projection models based on past five-year estimates. New UN requirements added the need to update the package so that analysts can conduct estimation of past TFR accounting for uncertainty, and make corresponding forecasts for both five-year and annual time periods.

To make probabilistic TFR projections accounting for past TFR uncertainty, the user will follow four steps in the following order:
\begin{enumerate}
	\item Data assembly (optional):
	\begin{enumerate}
		\item Prepare a dataset of raw TFR values. By default, the World Fertility Data 2019 \citep{UN2019WFD} is used.
		\item Assemble a dataset of reference (initial) TFR values for all countries and time periods. By default, the \textit{the World Population Prospects} \citep{UN2019} in the \pkg{wpp2019} package is used.
	\end{enumerate}
	\item Model estimation:
	\begin{enumerate}
		\item Train linear models to estimate systematic bias and standard deviation for each observation from the raw TFR dataset.
		\item Given the reference TFR, calculate the start period of Phase II and the start period of Phase III for each country ($\tau_c$ and $\lambda_c$).
		\item Run the MCMC process to obtain posterior samples of the Phase II and Phase III model parameters, and posterior samples of the past TFR for all countries.
	\end{enumerate}
	\item Generate future TFR trajectories as discussed in Section~\ref{sec:projections}. 
	\item Analyze the outputs using a set of functions that summarize, plot, diagnose and export the results of the three steps above.
\end{enumerate}

As described by \cite{vsevvcikova2011bayestfr}, steps 2 and 3 are relatively time-consuming. Adding the estimation uncertainty feature as well as working with annual estimates adds even more run time. Even though we optimized the code wherever possible, it takes several hours to complete these steps in a production-like setting. 
% Thus, we still kept the independence and durability of the results from them, including the parameters considered previously, as well as the past TFR estimates. 

The following sections describe the steps above in more detail, especially the parts that are different from \cite{vsevvcikova2011bayestfr}. We will elaborate on how to use the new features, as well as how to use the package in the original way. We will demonstrate the package on a realistic example with a relatively  large number of MCMC iterations, which might take several hours to process. Therefore, users who wish to explore the functionality quickly should reduce the number of iterations to the order of magnitude of 10. However note that since the Metropolis-Hastings step for the TFR updates will have an acceptance rate of around 30\%, a small number of iterations will result in estimation plots that are less smooth than what we will present in this article.

\subsection{Data assembly and estimation settings}\label{sec:rawdata}
The datasets assembled in this step will be passed to the main estimation function, \code{run.tfr.mcmc}, which now has additional arguments for this purpose. It can be specified what raw TFR data to use, whether to estimate and predict annually (logical argument \code{annual}), and whether to use the AR(1)  model in Phase II as defined  in Equation~\ref{model3} (logical argument \code{ar.phase2}). 

The argument \code{uncertainty = TRUE} specifies that uncertainty about the past is incorporated into the estimation.
In this case, a raw TFR dataset can be provided. By default, the World Fertility Data 2019 \citep{UN2019WFD} is used. This dataset contains 12,079 records for 201 countries, each of which includes the corresponding estimation method and data source. These are then used by the model as data quality covariates  in Equation~(\ref{model:unc}).

The default raw TFR dataset can be viewed via
\begin{CodeInput}
R> data("rawTFR")
R> head(rawTFR)
\end{CodeInput}
\begin{CodeOutput}
 country_code year  tfr   method source
1            4 1965 7.97 Indirect Census
2            4 1966 8.21 Indirect Census
3            4 1967 8.32 Indirect Census
4            4 1968 8.23 Indirect Census
5            4 1969 8.07 Indirect Census
6            4 1970 7.98 Indirect Census
 \end{CodeOutput}

The default covariates are \code{c("source", "method")}. Users can provide their own file and covariates of their choice. 
Required columns are ``country\_code'', ``year'' and ``tfr''.
The name of this file is passed to the argument \code{my.tfr.raw.file},  names of categorical variables to the argument \code{covariates}, and names of continuous variables to the argument \code{cont_covariates}.

An additional option allows an analyst to specify that vital registration data from selected countries are unbiased, if there is a belief that these data are not systematically biased in a particular direction. (Note that this is not the same as saying that these data are perfect.)  The UN country codes of those countries can be passed into the argument \code{iso.unbiased}. In this case, the bias and standard deviation of the records of those countries for which the \code{source} column specifies ``VR'' (as Vital Registration) are forced to be equal to 0 and  to be near 0, respectively (in practice the standard deviation is set to 0.0161). This option targets fine-tuning of the estimation of developed countries, especially because the annual TFR estimates are often not openly accessible.

The second dataset to assemble is a dataset on a reference, or initial, TFR. Its file name is passed into the argument \code{my.tfr.file}. If \code{uncertainty = FALSE}, this dataset is considered in the estimation as the true observed TFR. Otherwise, it is used as the starting points of TFR in the MCMC process. By default, if \code{my.tfr.file} is not given, the \code{tfr} dataset from the \pkg{wpp2019} package is used for this purpose, which is a five-year dataset. Thus, if \code{annual = TRUE}, a linear interpolation of the default dataset is computed. 

\subsection{Fitting the TFR model}
Most  arguments of \code{run.tfr.mcmc} remain the same as described in \cite{vsevvcikova2011bayestfr}. Importantly,  \code{start.year} and  \code{present.year} set the first and the last year of the time series included in the computation, respectively. The arguments \code{nr.chains}, \code{iters} and \code{output.dir} determine the number of chains, the number of iterations and the directory to store the MCMC simulated values, respectively.

In the prior version of  \pkg{bayesTFR}, the function \code{run.tfr.mcmc} was designed to obtain a posterior sample of Phase II model parameters. The estimation of Phase III parameters (as defined in Equation~\ref{modelphase3}) is implemented in the function \code{run.tfr3.mcmc}. When building a full probabilistic model as described in \cite{liu2018accounting}, the MCMC steps for updating TFR will affect both phases. Thus, if \code{uncertainty = TRUE}, the new \code{run.tfr.mcmc} function combines the estimation of both phases, and there is no need to invoke the \code{run.tfr3.mcmc} function explicitly. We call this method a ``one-step estimation". However, this is not the case if uncertainty about the past is not taken into account. In this case, the workflow of estimating Phase II and Phase III separately remains and is referred to as a a ``two-step estimation".

The various combinations of the possible settings of the arguments \code{annual} and \code{uncertainty} are summarized in Table~\ref{tbl0}. We have marked each cell with a letter which will be referred to in the remainder of this section. 
\begin{table}[!htb]
	\centering
	\begin{tabular}{c|c|c}
		\multirow{2}{*}{\bf annual} & \multicolumn{2}{c}{\bf uncertainty} \\ \cline{2-3} 
		&    \code{TRUE}       &       \code{FALSE}    \\ \hline
\multirow{3}{*}{\code{TRUE}}		&     {\bf A} & {\bf B} \\
& one-step estimation;      &    two-step estimation;       \\
& Phase II - AR(1) allowed & Phase II - AR(1) allowed \\ \hline
\multirow{2}{*}{\code{FALSE}} 		&      {\bf C}  & {\bf D}\\
& one-step estimation   &     two-step estimation
	\end{tabular}
\caption{\small Possible combinations in fitting TFR projection model}\label{tbl0}
\end{table}

As described in Section~\ref{sec:annual}, when using the annual model (cells A and B), adding the autoregressive component in Phase II as defined in Equation~(\ref{model3}) should be considered. The option is controlled via the logical argument \code{ar.phase2} which should be passed to the \code{run.tfr.mcmc} function. If \code{ar.phase2} is set to \code{TRUE} the MCMC process performs an extra slice sampling step for estimating $\phi$, an extra country-independent parameter in the model. The argument is ignored if \code{annual} is \code{FALSE}.

\subsubsection{Starting a new simulation with two-step estimation}\label{sec:not unc}
The two-step estimation should be performed if uncertainty about the past is not taken into account (cells B and D in Table~\ref{tbl0}).
The main differences between cells B and D are the setting of prior distributions as described in Section \ref{sec:annual}, and whether the autoregressive component can be included in the model. Here we give an example of a  simulation with annual data (cell B) without the autoregressive component. However, we will not use this example further in the text, as the main focus of the article is on cell A which will be discussed in the next section.

Our example simulation consists of three MCMC chains, each of which is 5000 iterations long where thinning is disabled. (As noted earlier, the user is advised to decrease the number of iterations to the order of ten for faster processing.) We will save the simulation results to a directory called ``annual''. 
\begin{CodeInput}
R> annual <- TRUE
R> nr.chains <- 3
R> total.iter <- 5000
R> thin <- 1
R> simu.dir <- file.path(getwd(), "annual")
\end{CodeInput}

The first step is to launch an estimation of Phase II:
\begin{CodeInput}
R> m2 <- run.tfr.mcmc(output.dir = simu.dir, nr.chains = nr.chains, 
+    iter = total.iter, thin = thin, annual = annual)
\end{CodeInput}

The second step is to start an estimation of Phase III:
\begin{CodeInput}
R> m3 <- run.tfr3.mcmc(sim.dir = simu.dir, nr.chains = nr.chains, 
+    iter = total.iter, thin = thin)
\end{CodeInput}

Here, we are using the same number of chains and iterations for Phase~II and Phase~III. However, this is not a requirement, but rather depends on the MCMC convergence. Even the $3\times 5000$ iterations might be not enough to reach convergence, but will usually give realistic outputs. Setting \code{total.iter = 62000} or \code{total.iter = "auto"} will most likely result in full convergence.

\subsubsection{Starting a new simulation with one-step estimation}
We now show an example of a simulation with uncertainty which is performed with one step only (cells A and C in Table \ref{tbl0}). In particular, here we set \code{annual} to \code{TRUE} (cell A), but the same function would be used if \code{annual} is \code{FALSE} (cell C). We will also include the Phase II - AR(1) model (\code{ar.phase2}) which would not have any effect in cell C. The results will be saved in the directory ``annual\_unc". We will use this simulation throughout the article. 
\begin{CodeInput}
R> annual <- TRUE
R> ar.phase2 <- TRUE
R> nr.chains <- 3
R> total.iter <- 5000
R> thin <- 1
R> simu.dir.unc <- file.path(getwd(), "annual_unc")
\end{CodeInput}
As in the previous case, this setting may not be enough to yield fully converged MCMC simulations, but will still give realistic outputs. The processing time is within a range of a couple of hours. For faster processing, set \code{total.iter = 50} for a toy simulation.  In addition, the \code{parallel} argument can be set to \code{TRUE}, in which case the three chains will be processed in parallel. In Section~\ref{sec:recommendation}, we will give recommendations for settings that yield fully converged MCMC simulations. When appropriate, we will use results from such converged simulations to present outputs. 

As mentioned in Section~\ref{sec:rawdata}, additional arguments of \code{run.tfr.mcmc} allow one to pass user-specific raw TFR data (\code{my.tfr.raw.file}), names of categorical covariates (\code{covariates}), names of continuous covariates (\code{cont_covariates}), or to specify countries with unbiased vital registration data (\code{iso.unbiased}). If the \code{iso.unbiased} option is used, the \code{covariates} argument should include the variable ``source'', or more specifically, the variable defined by the argument \code{source.col.name} which is ``source'' be default. In such a case, the function reduces the bias and standard deviation of records where the ``source'' column specifies ``VR''. In our example we will specify that the vital registration data of Canada and the USA (codes 124 and 840) are unbiased. 

To estimate both Phase II and Phase III, one could do
\begin{CodeInput}
R> m <- run.tfr.mcmc(output.dir = simu.dir.unc, nr.chains = nr.chains, 
+    iter = total.iter, annual = annual, thin = thin, 
+    uncertainty = TRUE, ar.phase2 = ar.phase2, 
+    iso.unbiased = c(124, 840))
\end{CodeInput}

In comparison to the two-step model, the training process here has an extra Metropolis-Hastings step per iteration for generating posterior TFR samples.

\subsubsection{Continuing an existing simulation}
If an existing simulation needs to be extended by more iterations, one would proceed as in the previous version of \pkg{bayesTFR}:
\begin{itemize}
\item Use \code{continue.tfr.mcmc} if the MCMCs were originally created via \code{run.tfr.mcmc}, regardless of whether one is in the one-step or the two-step estimation mode.
\item Use \code{continue.tfr3.mcmc} if the MCMCs were originally created via \code{run.tfr3.mcmc}.
\end{itemize}
Now suppose we want to extend the simulation in the previous section by 100 iterations. Then we could do
\begin{CodeInput}
R> m <- continue.tfr.mcmc(output.dir = simu.dir.unc, iter = 100)
\end{CodeInput}
(Set the \code{iter} argument to 10 if working with a toy simulation.) This will continue running both TFR phases in a one-step estimation while inheriting all settings from the original simulation, including  \code{uncertainty}, \code{annual}, \code{ar.phase2} or settings about the raw data. At the end of the processing, each chain will be 5,100 iterations long.

\subsection{Generating projections}
\label{sec:genproj}
The main function for generating projections is called \code{tfr.predict}. The new version of the package adds the argument \code{uncertainty}. If it is \code{TRUE} and the model was estimated taking uncertainty about the past into consideration, then that past uncertainty will be carried over to the projections. In this case, each future trajectory starts from a trajectory representing the past.

Suppose we want to generate projections represented by 1,000 posterior trajectories until the year 2100 based on the simulation stored in the directory given by \code{simu.dir.unc}, with burn-in of the first 2100 iterations for each chain.
This can be done using the following command:
\begin{CodeInput}
R> pred <- tfr.predict(sim.dir = simu.dir.unc, end.year = 2100,
+    burnin = 2100, nr.traj = 1000, uncertainty = TRUE)
\end{CodeInput}

The function takes the existing 5,100 iterations in each chain, removes the first 2,100 values and generates 1,000 TFR trajectories based on 1,000 equally spaced parameter values and past TFR, out of the remaining $3,000\times 3=9,000$ iterations. For a toy simulation, use \code{burnin = 20} and \code{nr.traj = 50}.
%Since the projection function will only generate TFR forecasts, the directory for storing these results are the same as described in Section 3.2 in \cite{vsevvcikova2011bayestfr}, and for each \code{rda} file storing all trajectories, it will include $3000\times 81$ values per file. Similarly, the thinned MCMC traces are also saved to disk in the sub-directory \code{"thinned\_mcmc\_}\textit{thin\_burnin"} in \code{"annual\_unc"} directory.

If \code{uncertainty} is set to \code{FALSE}, all future trajectories start from the last observed data point. If the estimation process accounted for uncertainty, but the projection does not, the starting value of the projections is the initial TFR value for the last observed time period. This is however not recommended but may serve the purpose of apples-to-apples comparisons.

%The way to access projections remain unchanged.

\subsection{Analyzing Results}
\label{sec:analresults}
If results are to be explored at a later time point, one can load the estimation object from disk using the command 
\begin{CodeInput}
R> m <- get.tfr.mcmc(simu.dir.unc)
\end{CodeInput}
For one-step estimation, this object contains information about both phases. For a two-step simulation, or if the Phase III object is to be extracted explicitly, use
\begin{CodeInput}
R> m3 <- get.tfr3.mcmc(simu.dir.unc)
\end{CodeInput}

Similarly, to load the prediction object from disk, do
\begin{CodeInput}
R> pred <- get.tfr.prediction(simu.dir.unc)
\end{CodeInput}

\subsubsection{Summary functions}
For the summary statistics of the estimation object in this section, we will use the following thinning and burn in settings:
\begin{CodeInput}
R> thin <- 3
R> burnin <- 2100
\end{CodeInput}
Use \code{thin <- 1} and \code{burnin <- 20} if you're working with the toy simulation.

To view a summary of country-independent parameters, one can use
\begin{CodeInput}
R> summary(m, thin = thin, burnin = burnin)
\end{CodeInput}
Since the object \code{m} was got by one-step estimation, the output includes estimation summaries for both phases. In comparison to the previous version of the package, Phase II contains one additional parameter, namely ``rho\_phase2'' which represents $\phi$ in model (\ref{model3}). As with any other parameter, the name, or multiple parameter names, can be passed to the function to view summary statistics  for those selected parameters.

\begin{CodeInput}
R> summary(m, par.names = c("rho_phase2", "sigma0"), 
+    thin = thin, burnin = burnin)
\end{CodeInput}

\begin{CodeOutput}
MCMCs of phase II
=================
Number of countries: 201
Hyperparameters estimated using 201 countries.
WPP: 2019
Input data: TFR for period 1950-2020
Time interval: annual

Iterations = 2103:5100
Thinning interval = 3 
Number of chains = 3 
Sample size per chain = 1000 

1. Empirical mean and standard deviation for each variable, 
   plus standard error of the mean:

         Mean      SD  Naive SE Time-series SE
rho_phase2 0.6988 0.04117 0.0007516       0.005133
sigma0     0.0562 0.01111 0.0002028       0.002038

2. Quantiles for each variable:

          2.5%     25%     50%     75%  97.5%
rho_phase2 0.59876 0.67408 0.71200 0.72904 0.7470
sigma0     0.04876 0.05162 0.05418 0.05694 0.0853
\end{CodeOutput}

The full list of parameter names for Phase II can be obtained via
\begin{CodeInput}
R> tfr.parameter.names(meta = m$meta)
\end{CodeInput}

\begin{CodeOutput}
[1] "alpha"  "alphat"  "delta"   "Triangle4"  "delta4"  "psi"  "chi"         
[8] "a_sd"  "b_sd"  "const_sd"  "S_sd"  "sigma0"  "mean_eps_tau"  "sd_eps_tau"  
[15] "rho_phase2"  
\end{CodeOutput}

Passing the \code{meta} argument is needed to identify that the simulation contains a Phase II - AR(1) estimation, and thus it contains the 
``rho\_phase2'' parameter. Phase III parameter names are not dependent on the simulation, thus no \code{meta} argument is needed: 
\begin{CodeInput}
R> tfr3.parameter.names()
\end{CodeInput}
\begin{CodeOutput}
[1] "mu"        "rho"       "sigma.mu"  "sigma.rho" "sigma.eps"
\end{CodeOutput}

Specifying a country in the \code{summary} function will show results for the country-specific parameters of that country. This is done via the \code{country} argument which accepts either the name or the numerical code of the country, as well as an ISO-2 or ISO-3 character code. This is the case for any function in the package that accepts the  the \code{country} argument, as is shown throughout the paper.  

As for the parameters to summarize, functions \code{tfr.parameter.names.cs()} and \\ \code{tfr3.parameter.names.cs()} list the allowed parameter names for Phase II and Phase III, respectively. For a simulation that took into account uncertainty about the past, there is an additional country-specific parameter, called ``tfr,'' capturing that uncertainty. It is not listed explicitly via the above functions, but it can be explored like any other parameter. For the \code{summary} function it means passing it to the \code{par.names.cs} argument. For example,  to view summary statistics of TFR estimation for Nigeria, we can do

\begin{CodeInput}
R> summary(m, country = "Nigeria", par.names.cs = "tfr",
+    thin = thin, burnin = burnin)
\end{CodeInput}
The tabular sections of the output contain one row per past observed period each (by default 71, i.e., from 1950 to 2020). To select a subset we can specify which time periods we are interested in as \code{tfr\_}\textit{time}. For example, to view results for time periods 1, 30 and 70 (corresponding to 1950, 1979 and 2019) we do
\begin{CodeInput}
R> summary(m, country = "Nigeria", 
+    par.names.cs = c("tfr_1", "tfr_30", "tfr_70"), 
+    thin = thin, burnin = burnin)
\end{CodeInput}

\begin{CodeOutput}
...
1. Empirical mean and standard deviation for each variable,
   plus standard error of the mean:

             Mean     SD Naive SE Time-series SE
tfr_1_c566  6.281 0.2390 0.004363       0.045831
tfr_30_c566 6.709 0.0762 0.001391       0.007952
tfr_70_c566 5.622 0.4785 0.008735       0.039332

2. Quantiles for each variable:

	         2.5%   25%   50%   75% 97.5%
tfr_1_c566  5.765 6.135 6.290 6.438 6.731
tfr_30_c566 6.577 6.655 6.704 6.760 6.879
tfr_70_c566 4.827 5.113 5.765 5.916 6.320
\end{CodeOutput}

\subsubsection{Exploring TFR estimation}
In addition to summary statistics, one can explore the estimated trajectories as well as any quantiles of the past TFR estimates.
For example,
\begin{CodeInput}
R> nigeria_obj <- get.tfr.estimation(country = "NG", 
+    sim.dir = simu.dir.unc, burnin = burnin, thin = thin, 
+    probs = c(0.025, 0.1, 0.5, 0.9, 0.975))
\end{CodeInput}
returns a list where trajectories are contained  in the  element \code{tfr_table}. The number of rows corresponds to the number of trajectories (here $3000 = 3[\text{chains}]\cdot(5100 - 2100)/3[\text{thin}]$, or $120 = 3(60 - 20)$ for the toy simulation), while columns correspond to time periods (here 71). 
\begin{CodeInput}
R> dim(nigeria_obj$tfr_table)
\end{CodeInput}

\begin{CodeOutput}
[1] 3000   71
\end{CodeOutput}
The quantiles are contained in the element \code{tfr_quantile}:
\begin{CodeInput}
R> head(nigeria_obj$tfr_quantile)
\end{CodeInput}

%{\it TODO: Replace with correct results from a 5100 simulation.}
\begin{CodeOutput}
        2.5%      10%      50%      90%    97.5% year
1: 5.764658 5.968237 6.289810 6.585244 6.730593 1950
2: 5.803398 5.979124 6.287563 6.563627 6.684089 1951
3: 5.810477 6.013333 6.308801 6.538703 6.702456 1952
4: 5.837490 6.004007 6.314881 6.536717 6.703307 1953
5: 5.862094 6.017064 6.317159 6.540123 6.712419 1954
6: 5.846718 6.027068 6.325334 6.545919 6.701965 1955
\end{CodeOutput}
This element is missing if the \code{probs} argument is not given.

\begin{figure}
	\centering
	\begin{tabular}{cc}
		\includegraphics[width=0.5\textwidth]{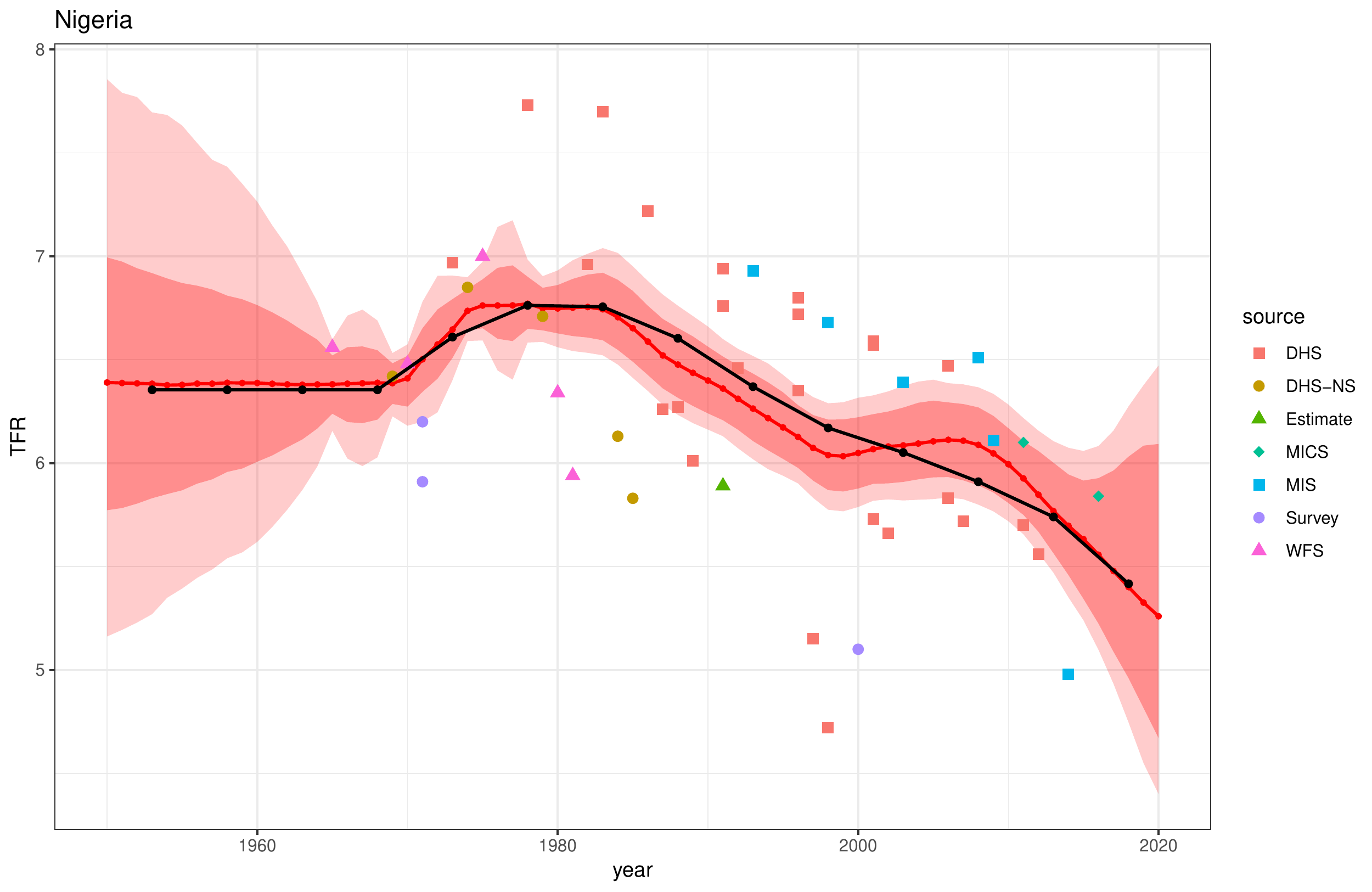} &
		\includegraphics[width=0.5\textwidth]{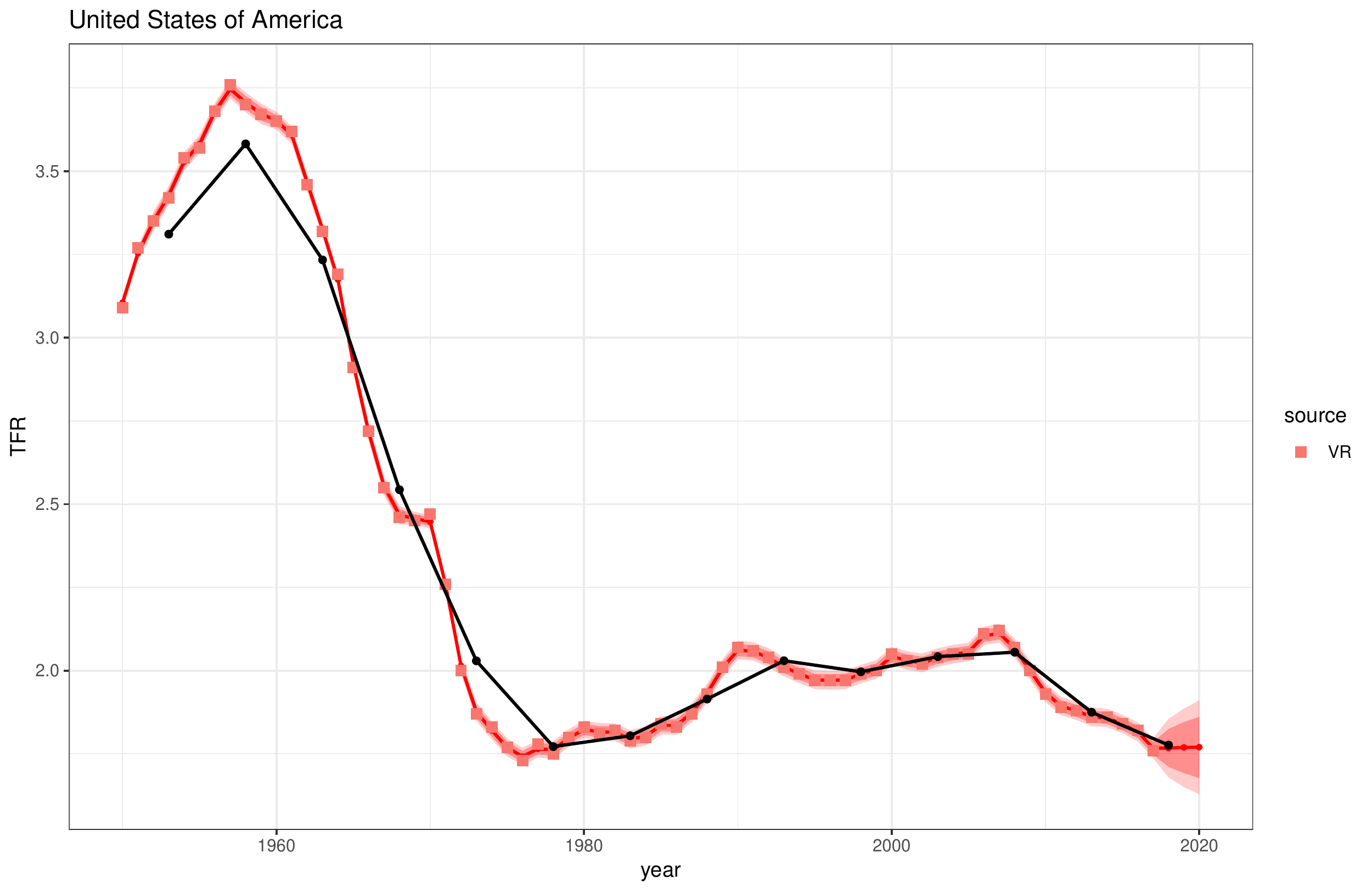}
	\end{tabular}
	\caption{\small Annual TFR estimation for Nigeria (left panel) and the United States (right panel), resulting from a converged simulation. The red line shows the posterior median, while the red shaded area shows the pointwise 80\% intervals, and the pink shaded areas shows the corresponding 95\% intervals.  The UN's 2019 WPP five-year estimates are shown by the black line.}
	\label{fig-estimation}
\end{figure}

For example, to plot the estimation with user-defined intervals, do:
\begin{CodeInput}
R> plot <- tfr.estimation.plot(country = 566, 
+    sim.dir = simu.dir.unc, burnin = burnin, thin = thin, 
+    pis = c(80, 95), plot.raw = TRUE)
R> print(plot)
\end{CodeInput}
The function uses the \pkg{ggplot2} package \citep{hadley2016ggplot} to visualize estimation uncertainty. Figure \ref{fig-estimation} shows results of the function call for Nigeria (as above) and the USA (\code{country = 840} or \code{country = "USA"}) using a converged simulation. 

Several arguments in this function need to be clarified: 
\begin{enumerate}
	\item \code{sim.dir}: Users can specify the location of the simulation set, or use the \code{mcmc.list} argument to pass the \code{m} object directly. For example \\\code{tfr.estimation.plot(mcmc.list = m, ...)}.
	\item \code{pis}: Specifies which probability intervals will be plotted. It is a vector of at most two elements.
	\item \code{plot.raw}: Logical argument which determines whether the raw data used for estimating past uncertainty are plotted. If \code{TRUE} and the estimation process was not based on the default data, it is recommended to provide the argument \code{grouping}, which should be one of the categorical covariates in the raw data set. This covariate defines the color and shape of the points, as seen in Figure \ref{fig-estimation} where the default grouping is the ``source'' column of the \code{rawTFR} dataset. 
	\item \code{save.image}: (not used in the call above) If  \code{TRUE}, the plot will be saved as a pdf file in the directory specified in the argument \code{plot.dir}, named ``tfr\_country\textit{code}.pdf''.
\end{enumerate}

\subsubsection{Exploring bias and standard deviation of observations}
Information about the bias and standard deviation of observations will give users an indication of the quality of the observations and whether these quantities were poorly estimated. 

Now suppose we are interested in the bias and standard deviation estimates of the observations for Nigeria. Then we could use
\begin{CodeInput}
R> bias_sd <- tfr.bias.sd(sim.dir = simu.dir.unc, country = 566)
\end{CodeInput}

The function will return a list with elements \code{model_bias},  \code{model_sd} and \code{table}. The \code{model_bias} and \code{model_sd} objects are of class \code{lm} and contain the linear models used to estimate the bias and standard deviation, respectively, while the \code{table} object includes the observed data points, data quality covariates, and the actual estimates for the specified country, here for Nigeria.
\begin{CodeInput}
R> summary(bias_sd$model_bias)
R> head(bias_sd$table)
\end{CodeInput}

The results are shown in Tables \ref{tbl4} and \ref{tbl5}. 
% latex table generated in R 3.6.0 by xtable 1.8-4 package
% Thu Jun 25 20:37:51 2020
\begin{table}[ht]
\centering
\begin{tabular}{rrrrr}
  \hline
 & Estimate & Std. Error & t value & Pr($>$$|$t$|$) \\ 
  \hline
(Intercept) & -0.43 & 0.10 & -4.37 & 0.00 \\ 
  covariate\_1DHS-NS & -0.31 & 0.17 & -1.76 & 0.09 \\ 
  covariate\_1Estimate & -0.14 & 0.37 & -0.39 & 0.70 \\ 
  covariate\_1MICS & 0.72 & 0.27 & 2.68 & 0.01 \\ 
  covariate\_1MIS & 0.29 & 0.16 & 1.79 & 0.08 \\ 
  covariate\_1Survey & -0.21 & 0.23 & -0.94 & 0.35 \\ 
  covariate\_1WFS & -0.18 & 0.17 & -1.06 & 0.30 \\ 
  covariate\_2Indirect & 0.80 & 0.11 & 7.05 & 0.00 \\ 
   \hline
\end{tabular}
\caption{\small Linear model for bias obtained from \code{summary(bias\_sd\$model\_bias)} for Nigeria.}\label{tbl4}
\end{table}

% latex table generated in R 3.6.0 by xtable 1.8-4 package
% Thu Jun 25 20:45:28 2020
\begin{table}[ht]
\centering
\begin{tabular}{llrr}
  \hline
method & source & bias & std\\ 
  \hline
Indirect & WFS & 0.18 & 0.13 \\ 
  Indirect & DHS-NS & 0.06 & 0.09 \\ 
  Direct & Survey & -0.64 & 0.64 \\ 
  Indirect & DHS & 0.37 & 0.28 \\ 
  Direct & WFS & -0.61 & 0.61 \\ 
  Direct & DHS-NS & -0.74 & 0.74 \\ 
   \hline
\end{tabular}
\caption{\small Bias and standard deviation of each observation obtained from \code{bias\_sd\$table} for Nigeria.}\label{tbl5}
\end{table}

To generate the estimates in the \code{table} object, the \code{predict} S3 method is applied to the \code{model_*} objects. Then the following steps are performed:
\begin{enumerate}
	\item For some countries, the number of data points is very small for several groups. This could lead to a large bias, but a very small variance. As a result, the estimation will be unreasonably concentrated on the bias-adjusted data points. In this case, the standard deviation estimates were adjusted as $\max(0.1, |\text{bias}|/2)$. The reason for such an adjustment is that it is unlikely that one observation is very biased but with a very small relative standard deviation. 
It is also unlikely that there is a source of data that is very precise (with standard deviation less than 0.1), but is only collected once.
	\item For countries included in \code{iso.unbiased}, the model estimates are overwritten with zero or close to zero values as explained in Section~\ref{sec:rawdata}.
	\item Duplicates are dropped so that the combinations of data quality covariates are unique.
\end{enumerate}

The output can help to detect problematic estimates on certain data points so that adjustments can be made by the analyst if necessary. In the example above, the estimated bias and standard deviation for the Indirect method and nationwide DHS surveys were 0.06 and 0.09, respectively. These estimates were derived based on only three data points in this category, and all of them were very close to the UN estimates (three of the brown dots in Figure \ref{fig-estimation} in year 1969, 1974 and 1979). Since the number of data points from nationwide DHS estimates is small (3 data points), the estimated bias (0.06) and standard deviation (0.09) may be too small. 

\subsubsection{Exploring TFR prediction}
Plotting the posterior sample of projected TFR trajectories is done via the  \code{tfr.trajectories.plot} function. The updated version of the package incorporates uncertainty about the past, if taken into account in the estimation and projection. For example, to plot the prediction of TFR for Burkina Faso contained in the \code{pred} object created in Section~\ref{sec:genproj} or at the beginning of Section~\ref{sec:analresults}, use
\begin{CodeInput}
R> tfr.trajectories.plot(pred, country = "Burkina Faso", nr.traj = 20, 
+    pi = c(80, 95), uncertainty = TRUE)
R> tfr.trajectories.plot(pred, country = "Burkina Faso", nr.traj = 20, 
+    pi = c(80, 95), uncertainty = FALSE)
\end{CodeInput}
Here, the parameter \code{uncertainty} is used to specify whether the uncertainty about the past TFR should be plotted together with the prediction. If \code{uncertainty} is \code{TRUE}, the optional parameters \code{thin, burnin, col\_unc} can be used to define the burn-in, thinning and the color for the past uncertainty plot. 

\begin{figure}
	\centering
	\begin{tabular}{cc}
		\includegraphics[width=0.5\textwidth]{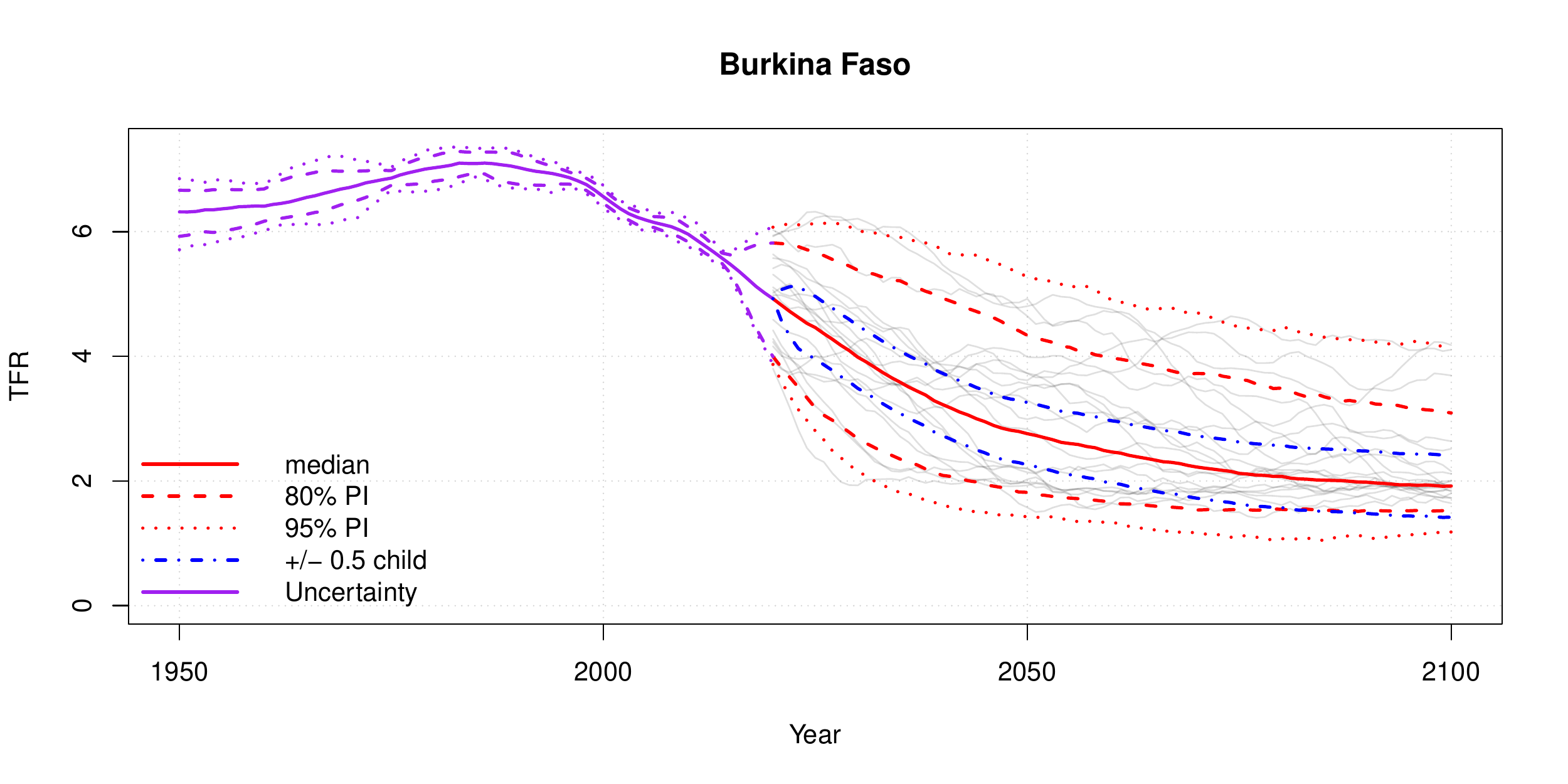} &
		\includegraphics[width=0.5\textwidth]{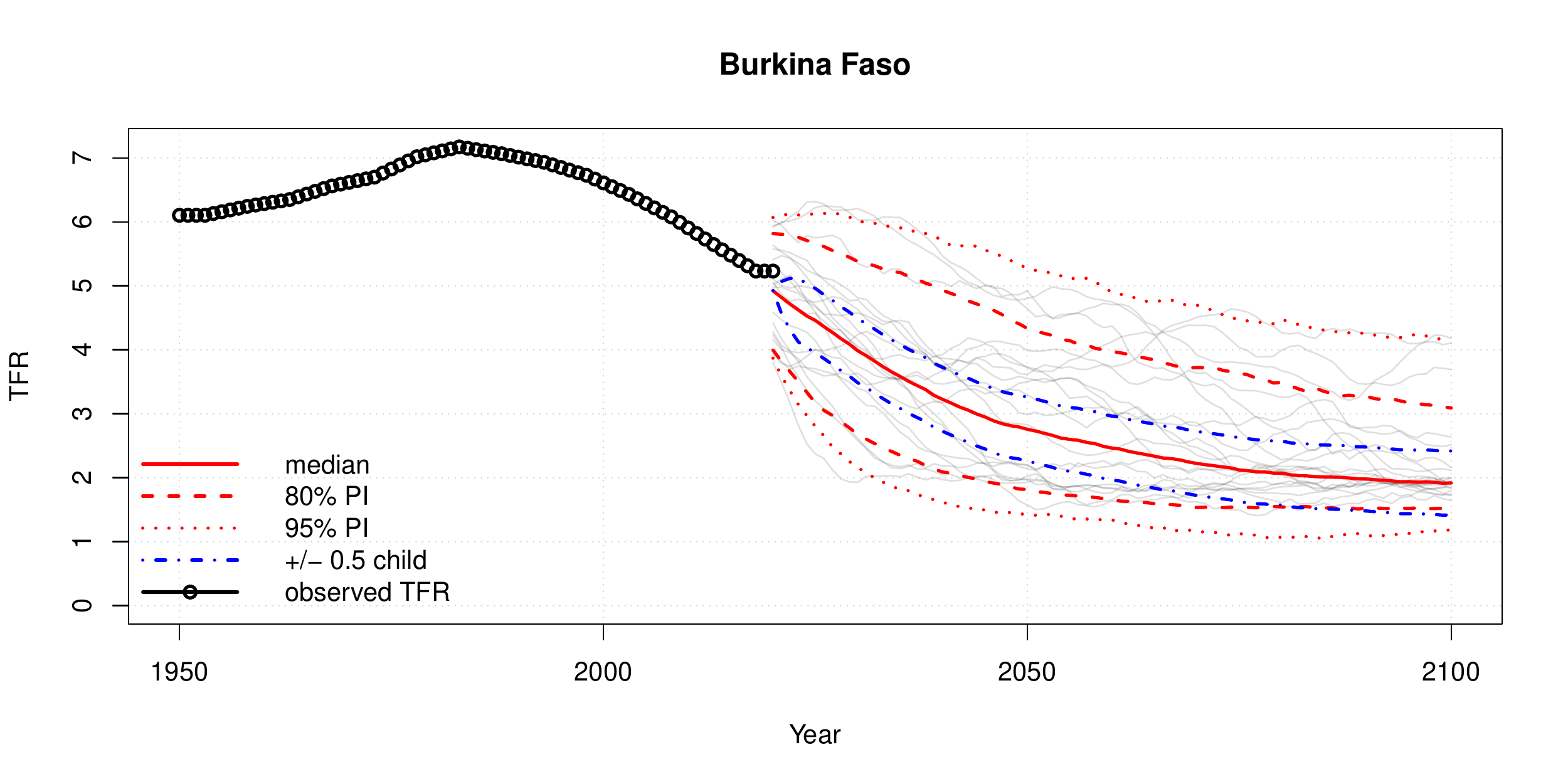}
	\end{tabular}
	\caption{\small TFR prediction (from a converged simulation) for Burkina Faso 
		with uncertainty about the past TFR (left panel) and without it (right panel). The black dots in the right panel represent the TFR used for initializing the simulation. In both panels, the red curves (solid, dashed and dotted) show the probabilistic prediction (median, 80\% and 95\% probability intervals), while the blue lines show the traditional UN scenarios of adding and removing a half a child to/from the main projection, here the median TFR. }
	\label{fig-prediction}
\end{figure}

The code above applied to a converged simulation results in the plots shown in Figure~\ref{fig-prediction}. 
If the user selects \code{uncertainty = FALSE} for a simulation where past uncertainty was taken into account (similarly to the right panel of Figure~\ref{fig-prediction}), the past TFR used for the initialization of the model is shown as the observed TFR. In this case, there could be a discontinuity between the last observed and the first projected time period.

These new arguments are also accepted by the \code{tfr.trajectories.plot.all} function which generates projection plots for all countries at once, as described by \cite{vsevvcikova2011bayestfr}.

TFR predictions in a tabular format can be explored using the \code{tfr.trajectories.table} and \code{summary} functions which work the same way as in the previous versions of the package, except that in the former case, the output now contains uncertainty information about the past. 

\begin{CodeInput}
R> tfr.trajectories.table(pred, country = "Burkina Faso")
\end{CodeInput}

%{\it TODO: Replace with correct results from a 5100 simulation.}
\begin{CodeOutput}
       median    0.025      0.1      0.9    0.975 -0.5child +0.5child
1950 6.260554 5.796712 5.950538 6.628323 6.864889        NA        NA
1951 6.247358 5.836737 5.966940 6.608714 6.815591        NA        NA
1952 6.263510 5.837363 5.983308 6.601735 6.812113        NA        NA
1953 6.272365 5.876240 6.002074 6.602380 6.806366        NA        NA
1954 6.284774 5.874250 6.006758 6.621279 6.836863        NA        NA
...
2095 1.971179 1.273660 1.555806 3.469681 4.344101  1.471179  2.471179
2096 1.968939 1.293491 1.551197 3.458955 4.271590  1.468939  2.468939
2097 1.974058 1.307082 1.554862 3.372069 4.239252  1.474058  2.474058
2098 1.961362 1.334648 1.552755 3.367219 4.246914  1.461362  2.461362
2099 1.951146 1.324320 1.552732 3.366420 4.272629  1.451146  2.451146
2100 1.948056 1.337980 1.549511 3.329382 4.228294  1.448056  2.448056
\end{CodeOutput}

\begin{CodeInput}
R> summary(pred, country = "Burkina Faso")
\end{CodeInput}

\begin{CodeOutput}
Projections: 80 ( 2021 - 2100 )
Trajectories: 1000
Phase II burnin: 2100
Phase II thin: 9
Phase III burnin: 2100
Phase III thin: 9

Country: Burkina Faso 

Projected TFR:
 	mean    SD 2.5%   5%  10%  25%  50%  75%  90%  95% 97.5%
2020 5.00 0.663 3.94 4.01 4.15 4.44 4.98 5.61 5.94 6.04  6.19
2021 4.90 0.741 3.71 3.79 3.96 4.28 4.89 5.56 5.94 6.07  6.21
2022 4.81 0.809 3.46 3.60 3.77 4.13 4.80 5.51 5.92 6.09  6.26
2023 4.71 0.869 3.22 3.40 3.59 4.00 4.69 5.44 5.90 6.13  6.31
...
\end{CodeOutput}

\subsubsection{Exploring double logistic function}
\begin{figure}
	\centering
	\begin{tabular}{cc}
		\includegraphics[width=0.5\textwidth]{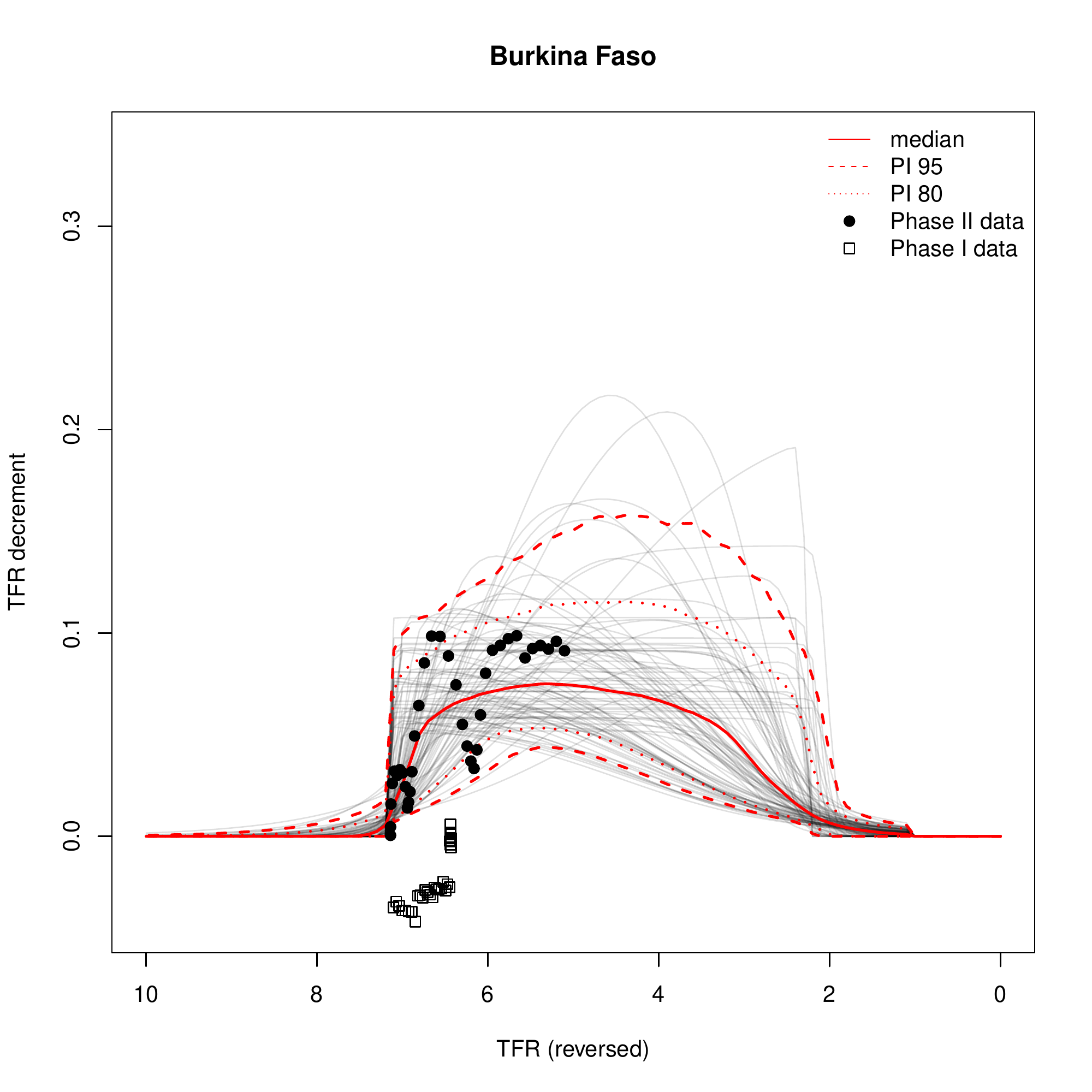} &
		\includegraphics[width=0.5\textwidth]{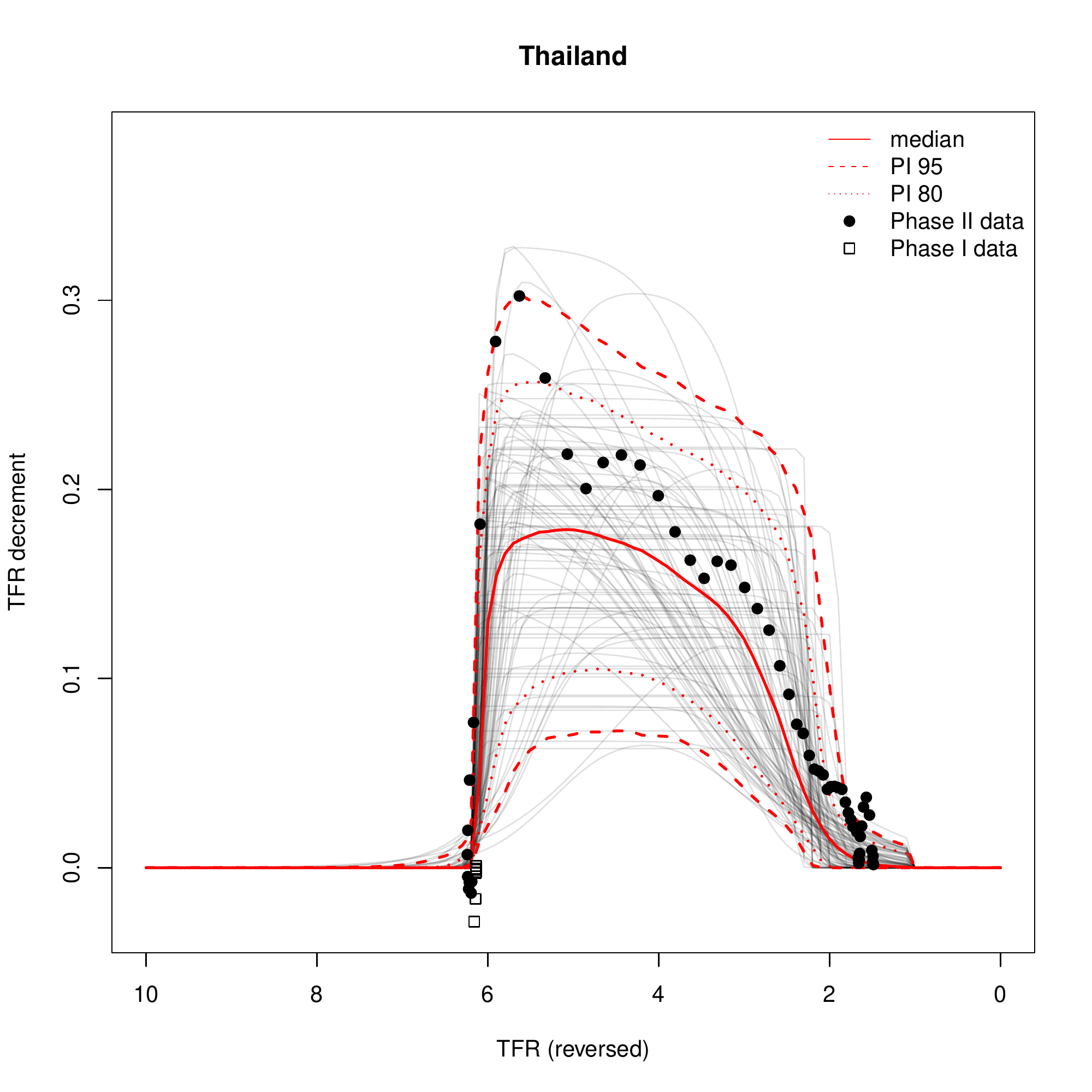}
	\end{tabular}
	\caption{\small Estimated Double Logistic curves (from a converged simulation) for 
		Burkina Faso (left panel) and Thailand (right panel). 
The data points (black dots and squares) are the estimated median decrements per year.}
	\label{fig-dlcurve}
\end{figure}

The double logistic function defined in (\ref{dlcurve}) can be viewed using 
\begin{CodeInput}
R> DLcurve.plot(country = "BFA", mcmc.list = m, 
+    burnin = burnin, pi = c(95, 80), nr.curves = 100)
\end{CodeInput}
Results can be seen in the left panel of Figure~\ref{fig-dlcurve}, while the right panel shows the result of the same call with \code{country = "Thailand"}. 

If a simulation contains information about past uncertainty, then the Phase II and I data (black dots and squares) represent decrements of the estimated TFR median. In case of an annual simulation, these are annual decrements, otherwise they would correspond to five-year decrements. If the projections were produced without taking past uncertainty into account, then the data points represent the observed decrements. 

This also applies to the \code{DLcurve.plot.all} function which plots the double logistic curves for all countries at once.

\subsubsection{MCMC traces, density and diagnosis}
To explore traces of the MCMC parameters, the existing functions \code{tfr.partraces.plot} (for country-independent parameters) and \code{tfr.partraces.cs.plot} (for country-specific parameters) can be used. Similarly, for density plots, \code{tfr.pardensity.plot} and \code{tfr.pardensity.cs.plot} are available. 

As mentioned previously, there are two additional parameters in this version of the package, namely ``rho\_phase2", which is country-independent and defined in model (\ref{model3}), and ``tfr" which is a country-specific parameter. These two parameters can be used within the aforementioned functions, like any other parameters . 

\begin{figure}
	\centering
	\begin{tabular}{cc}
		\includegraphics[width=0.5\textwidth]{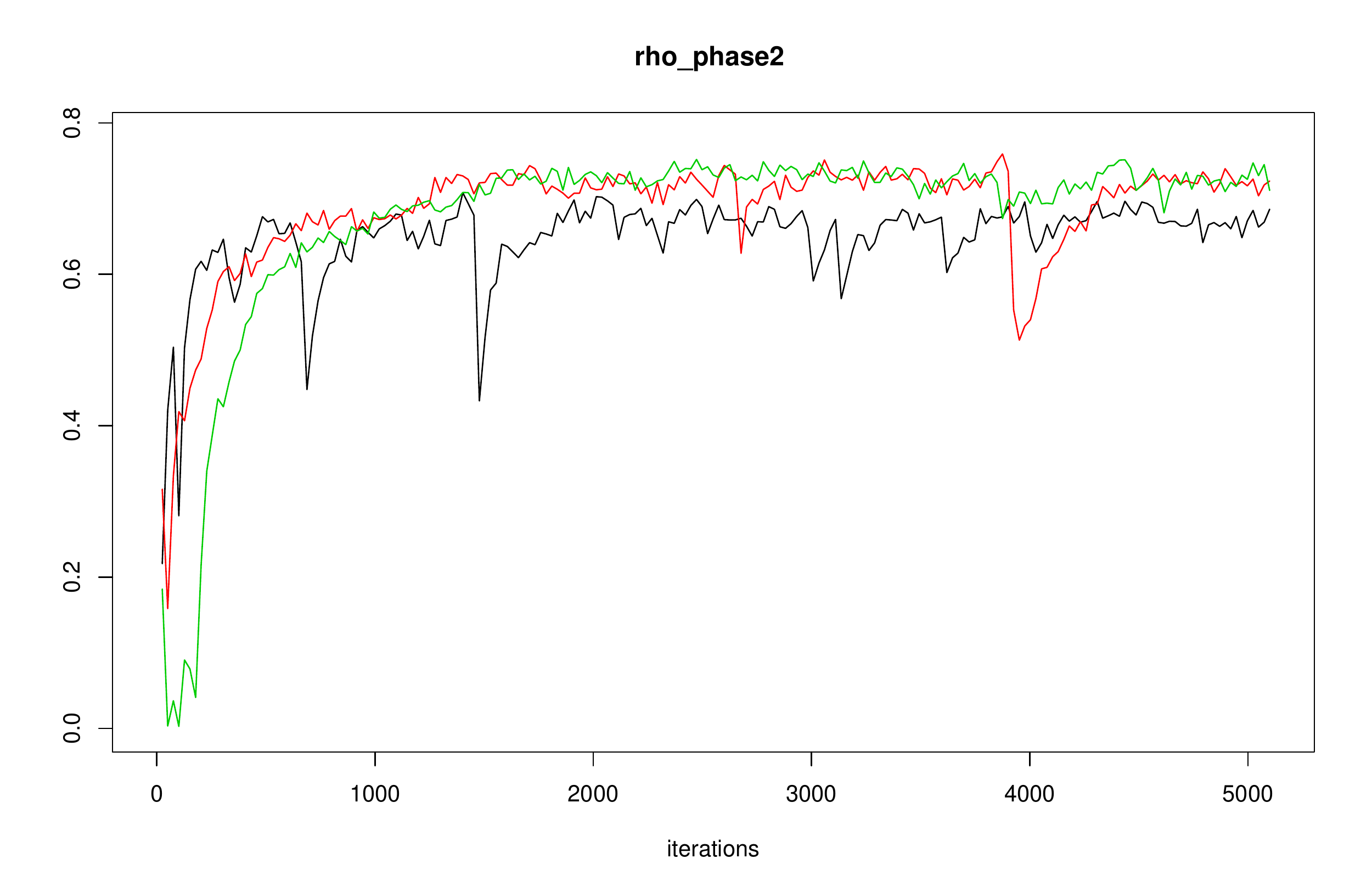} &
		\includegraphics[width=0.5\textwidth]{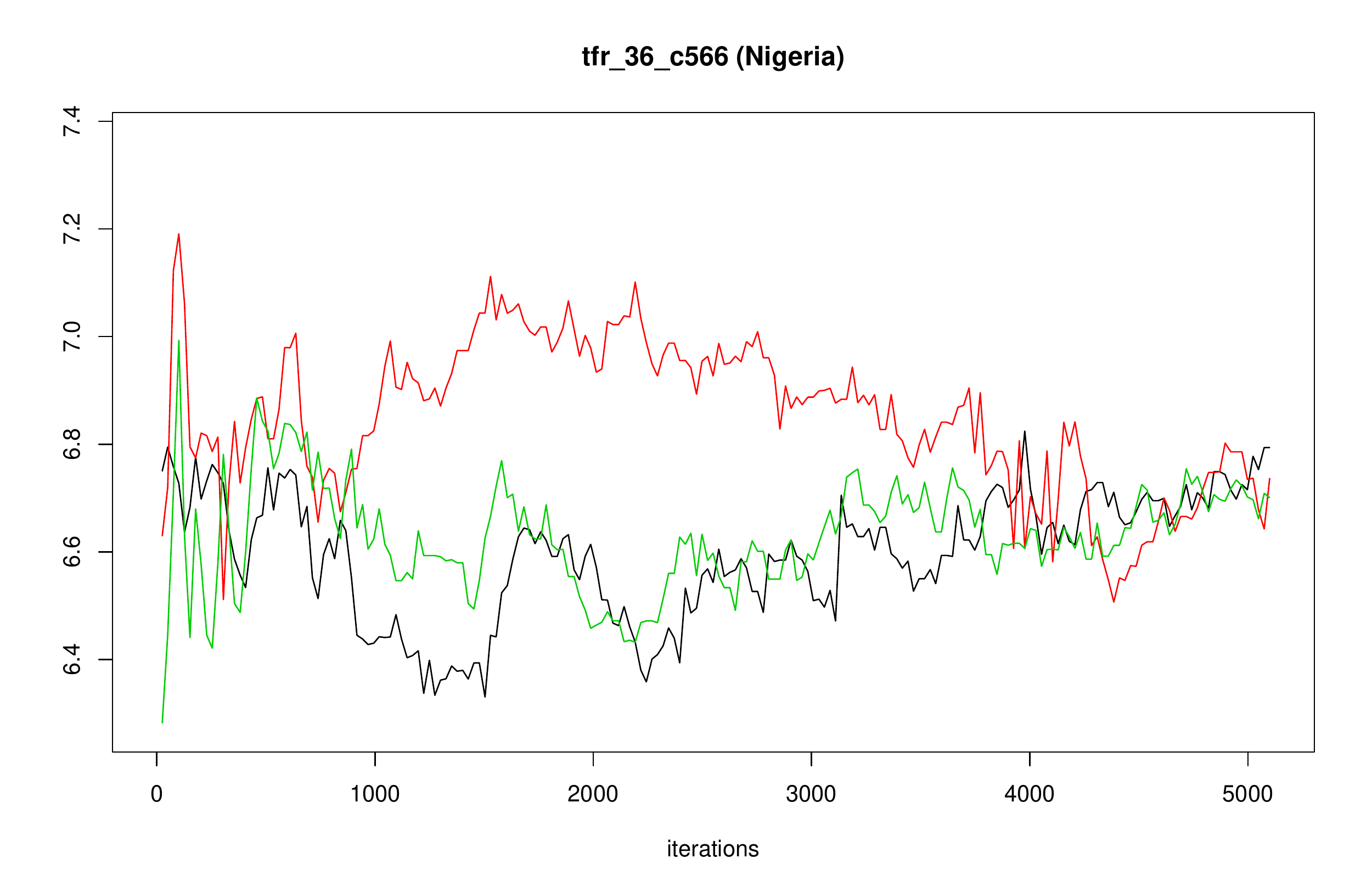}
	\end{tabular}
	\caption{\small Trace plots for  $\phi$ (left panel) and TFR of Nigeria in 1985 (right panel). }
	\label{fig-trace}
\end{figure}

\begin{figure}
	\centering
	\begin{tabular}{cc}
		\includegraphics[width=0.5\textwidth]{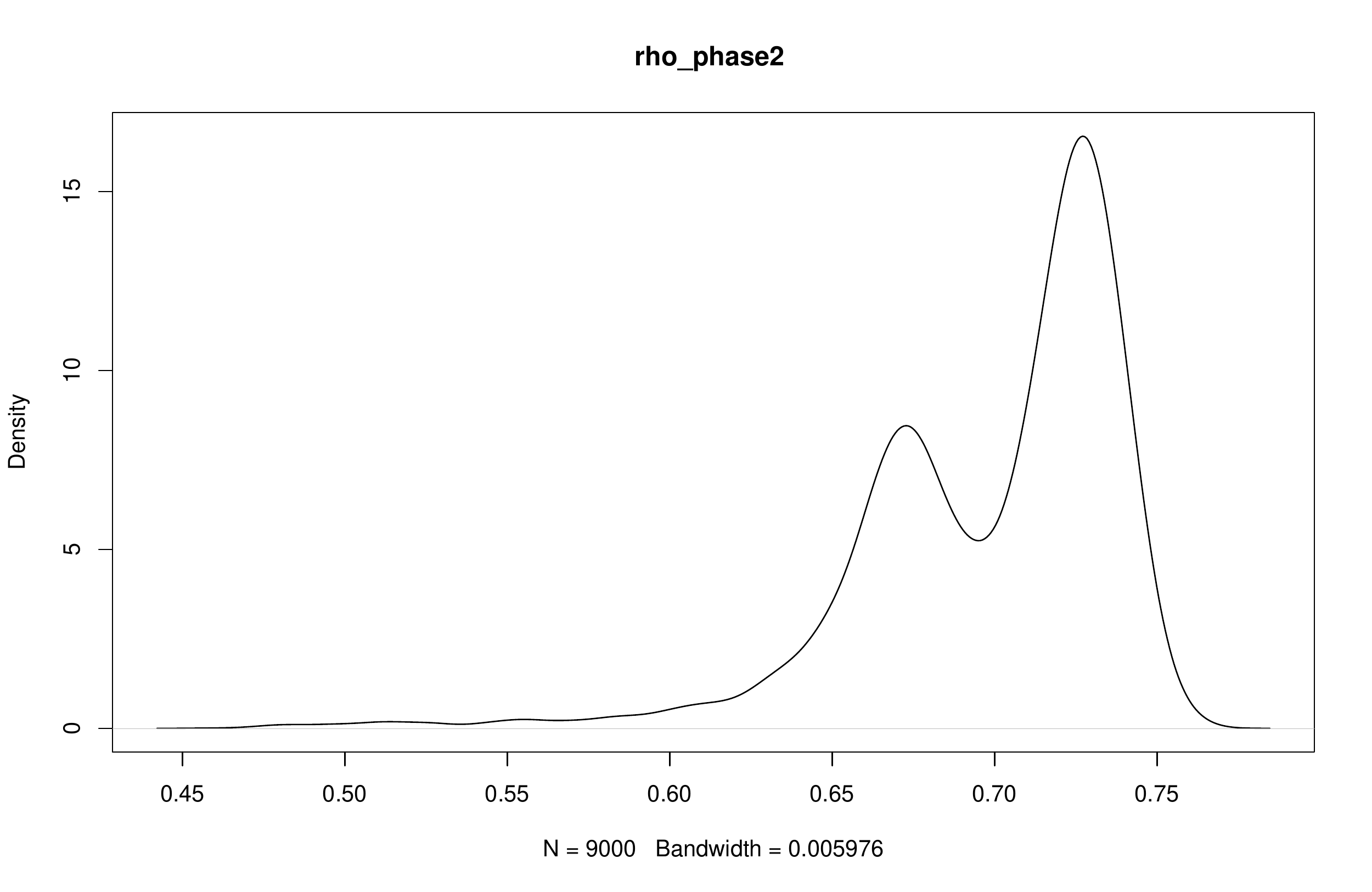} &
		\includegraphics[width=0.5\textwidth]{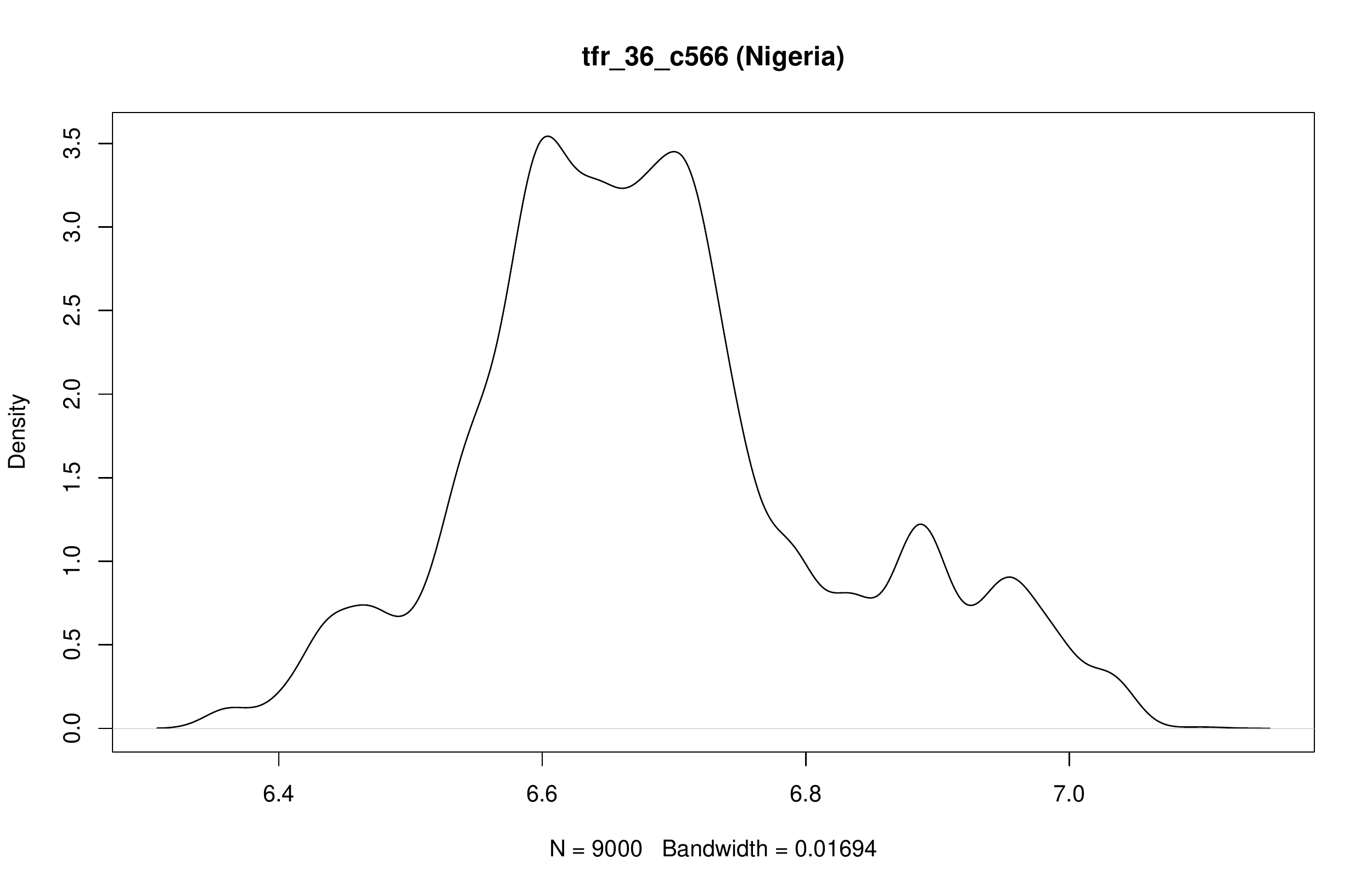}
	\end{tabular}
	\caption{\small Density plots for $\phi$ (left panel) and TFR of Nigeria in 1985 (right panel). }
	\label{fig-density}
\end{figure}

For example, the trace plots and the density plots of $\phi$ and Nigeria's TFR estimate in year 1985 (as shown in Figures \ref{fig-trace} and \ref{fig-density} ) can be visualized via
\begin{CodeInput}
R> tfr.partraces.plot(m, par.names = "rho_phase2", nr.points = 200)
R> tfr.partraces.cs.plot(m, country = "Nigeria", 
+    par.names = "tfr_36", nr.points = 200)
R> tfr.pardensity.plot(m, par.names = "rho_phase2", burnin = burnin)
R> tfr.pardensity.cs.plot(m, country = "Nigeria", 
+    par.names = "tfr_36", burnin = burnin)
\end{CodeInput}

To check if the MCMC algorithm has converged and adequately explored the parameter space,  the \code{tfr.diagnose} function can be used; see \cite{vsevvcikova2011bayestfr} for more details. In the case of one-step estimation, the function checks parameters from Phase II as well as Phase III. In the case of two-step estimation, one would use \code{tfr.diagnose} for assessing the convergence of Phase II parameters, and \code{tfr3.diagnose}  for assessing the convergence of Phase III parameters. Both functions accept a logical argument \code{express} which can disable or reduce the checking of country-specific parameters in order to speed up the process.

If the estimation includes uncertainty about the past, the assessment of country-specific parameters include the ``tfr'' parameter for each country and time period, in our case more than 14200 ``tfr'' parameters. In practice, it is often impossible to achieve convergence for all of them. Thus, we introduced the rule of accepting the ``tfr'' parameters as having converged if 95\% of them have converged.  

To apply the convergence diagnostics to our simulation, one could do
\begin{CodeInput}
R> tfr.diagnose(simu.dir.unc, thin = thin, 
+    burnin = burnin, express = TRUE)
\end{CodeInput}

As mentioned earlier, in our illustrative code examples the MCMC algorithm has  not been run for long enough to achieve full convergence.  See Section~\ref{sec:recommendation} for alternative settings. Note that the toy simulation we proposed earlier cannot be checked for convergence, as there is a requirement of a minimum number of iterations per chain, which the toy simulation does not satisfy.

\subsection{Estimating a small set of countries}
The Bayesian framework we have shown so far is designed to estimate all countries of the world at once, where the historical experience of an individual country influences the distribution of its own parameters as well as of the world parameters, while using the same settings for all countries. However, this is not always practical for several reasons:
\begin{enumerate}
\item Analysts might want to experiment with settings for individual countries without waiting several hours for a simulation of the whole world to finish. 
\item Different sets of covariates might be needed to estimate different countries. 
\item Countries with unusual historical patterns or very small countries might be excluded from the simulation in order not to bias the world parameters. 
\end{enumerate}

It was the last reason, as well as the need to include aggregations in the estimation, that motivated us to implement the \code{run.tfr.mcmc.extra} function in the original version of the package. The idea is that, while \code{run.tfr.mcmc} updates all parameters, the \code{run.tfr.mcmc.extra} function updates only the country-specific parameters of the specified countries, while re-using the existing distribution of the global parameters. 

Since the function was designed for special cases of countries or aggregations, the original implementation allowed the user to process only the locations that had not been included in the world simulation. With the  two additional use cases above, we have now relaxed that restriction and made it possible to rerun  and overwrite existing estimations of country-specific parameters and past TFR estimates for individual countries, while allowing the user to change various estimation settings. However, several global settings are not subject to change, such as 
switching between annual and five-year estimation, or changing the \code{ar.phase2} argument. 

Suppose that after running the simulation with the default data from the World Fertility Data, the user wishes to experiment with their own data that exclude Nigeria's questionable data points, such as the Indirect DHS-NS data points identified in Table~\ref{tbl5} as having unreasonably low standard deviations and biases. Unlike in the main simulation, the experiment will not force the VR data of the United States to have zero bias and variance. For that purpose, we will extract data for Nigeria (code 566) and the USA (code 840) from the default raw dataset discussed in Section~\ref{sec:rawdata}, remove the Indirect DHS-NS points for Nigeria and store them into a file called ``raw\_tfr\_user.csv'':
\begin{CodeInput}
R> countries <- c(566, 840)
R> myrawTFR <- subset(rawTFR, country_code %in% countries)
R> myrawTFR <- subset(myrawTFR, !(country_code == 566 
+    & method == "Indirect" & source == "DHS-NS"))
R> write.csv(myrawTFR, file = "raw_tfr_user.csv", row.names = FALSE)
\end{CodeInput}

For experimentation with the \code{run.tfr.mcmc.extra} function, we recommend copying the main simulation into a different directory and applying the function to the copy. This is because the processing overwrites the existing estimation results, and thus there is no way back to the original results in case the experiments do not yield satisfactory outputs. Here we will append ``\_extra'' to the directory name stored in \code{simu.dir.unc} and copy the content from \code{simu.dir.unc} into it. This step is equivalent to the command \code{"cp -r annual_unc annual_unc_extra"} on unix-based systems:
\begin{CodeInput}
R> simu.dir.extra <- paste0(simu.dir.unc, "_extra")
R> dir.create(simu.dir.extra)
R> file.copy(list.files(simu.dir.unc, full.names = TRUE), 
+    simu.dir.extra, recursive = TRUE)
\end{CodeInput}

To run the new estimation for the two selected countries, we can do
\begin{CodeInput}
R> run.tfr.mcmc.extra(sim.dir = simu.dir.extra, countries = countries, 
+    iter = total.iter, burnin = burnin, uncertainty = TRUE, 
+    my.tfr.raw.file = "raw_tfr_user.csv", 
+    covariates = c("source", "method"))
\end{CodeInput}
We recommend using the same values of \code{total.iter} and \code{burnin} as in the main simulation. 

\begin{figure}
	\centering
	\begin{tabular}{cc}
		\includegraphics[width=0.5\textwidth]{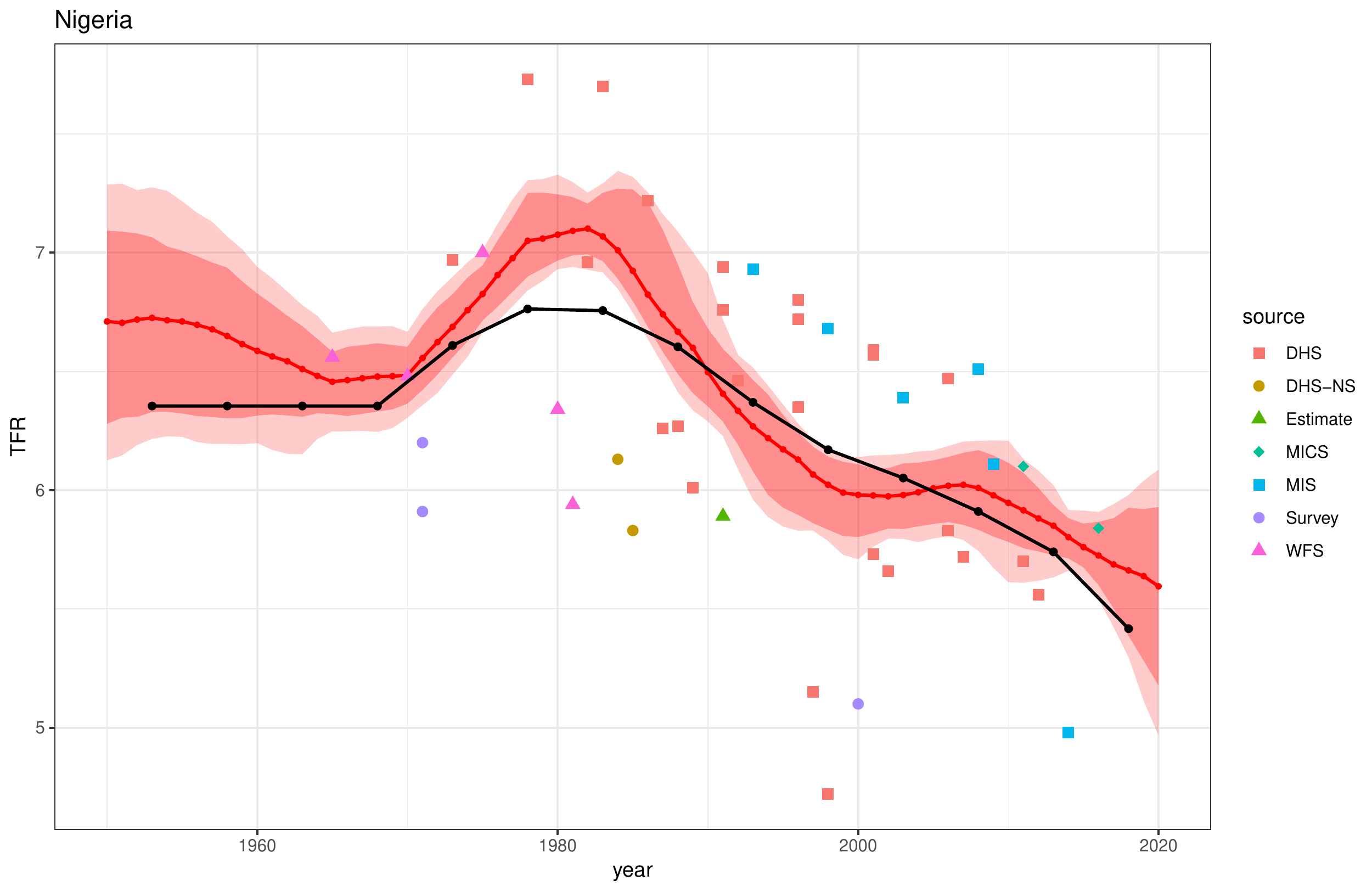} &
		\includegraphics[width=0.5\textwidth]{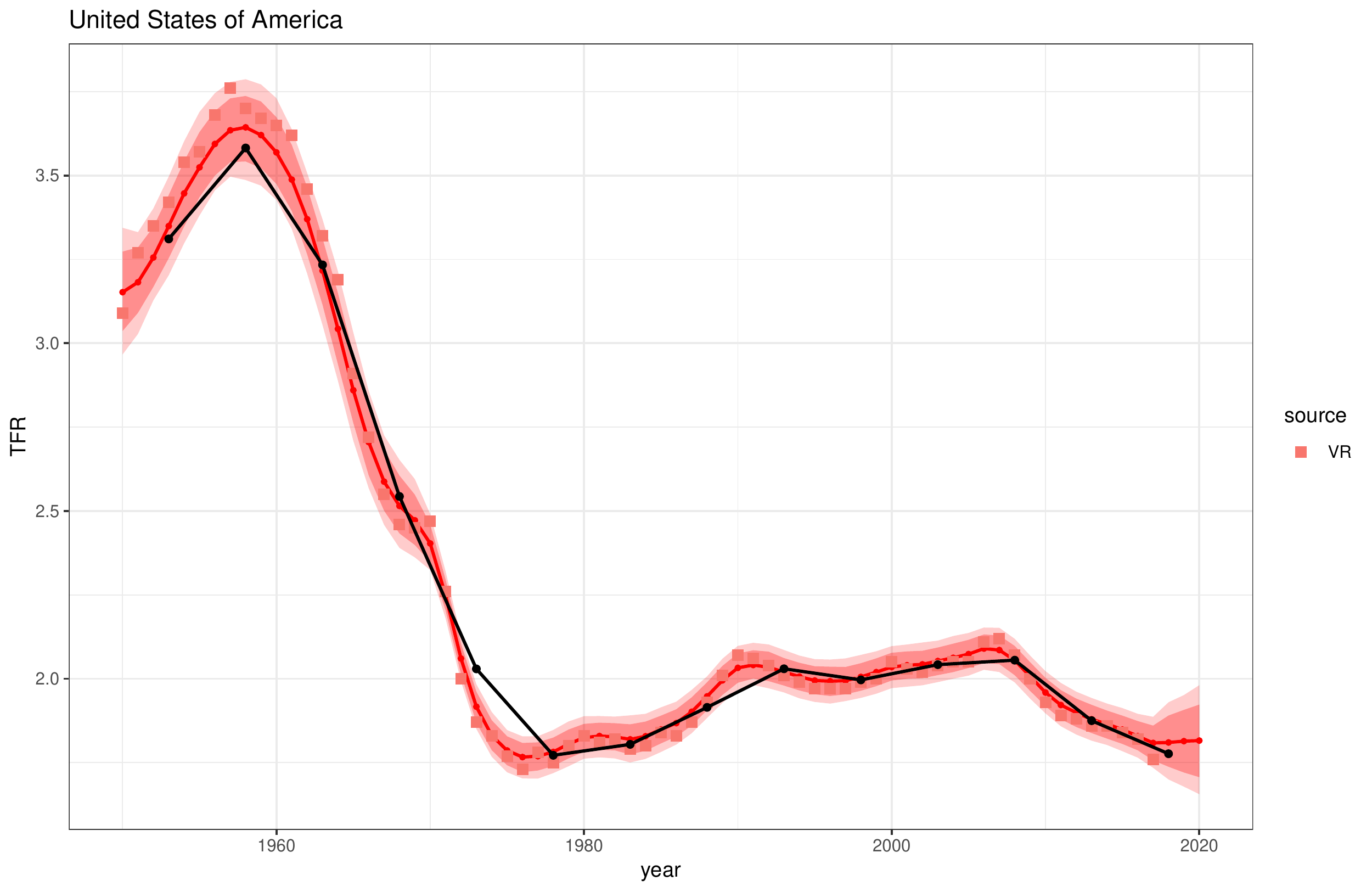}
	\end{tabular}
	\caption{\small TFR estimation for Nigeria (left panel) and the United States (right panel), resulting from a non-converged simulation with modified data set.}
	\label{fig-estimation-extra}
\end{figure}

To compare the new estimation results to those shown in Figure~\ref{fig-estimation} we again use the \\
\code{tfr.estimation.plot} function, now passing \code{simu.dir.extra} into the \code{sim.dir} argument. It can be seen in the left panel of Figure~\ref{fig-estimation-extra}  that excluding the Indirect DHS-NS data points for Nigeria changed the estimates, especially for 1979. The uncertainty increased for the United States (right panel of Figure~\ref{fig-estimation-extra}), since it was removed from the \code{iso.unbiased} set. 

Finally, the option \code{uncertainty = TRUE} can be used even in two-step estimation where uncertainty about the past was not taken into account. This is possible because we do not expect the global parameters to be significantly different in the two situations (i.e., with and without uncertainty).

\subsection{Structure of the output directory}\label{sec:output}
Having a look at the simulation directory, here ``annual\_unc'', one should see a structure similar to the following:
% This was created by
% tree annual_unc -P bayes*meta*
\begin{CodeOutput}
annual_unc
├── bayesTFR.mcmc.meta.rda
├── diagnostics
├── mc1
├── mc2
├── mc3
├── phaseIII
│   ├── bayesTFR.mcmc.meta.rda
│   ├── mc1
│   ├── mc2
│   └── mc3
├── predictions
└── thinned_mcmc_9_2100
    ├── bayesTFR.mcmc.meta.rda
    └── mc1
\end{CodeOutput}

The directories ``mc1'', ``mc2'' and ``mc3'' on the first level are generated by the \code{run.tfr.mcmc} function and contain results from the three chains of the Phase II estimation. Each of the directories contains one text file per parameter. The names of the hyperparameters and their corresponding notation are the same as described in Table 1 in \cite{vsevvcikova2011bayestfr}. In addition, the parameter ``rho\_phase2'' representing $\phi$ from Equation~\ref{model3} is also stored as a hyperparameter if the Phase~II-AR(1) is considered. The names of the files storing country-independent parameters consist of the parameter name and the suffix ``.txt'', while in the case of the files storing country-specific parameters the parameter name is followed by the suffix ``\_country\textit{code}.txt''. 

If uncertainty is taken into account, the MCMC algorithm also generates estimates for the past TFR data. These samples are considered as country-specific parameters, called ``tfr'', and thus stored in files  ``tfr\_country\textit{code}.txt''. They contain matrices of size the number of (thinned) iteration times the number of time periods. In the example above, the default starting year is 1950, and the present year is 2020, i.e., 71 years. Therefore, each file contains TFR estimates in 5100 rows  and 71 columns.

The file ``bayesTFR.mcmc.meta.rda'' on the first level stores meta information about the Phase~II estimation, which is contained in the \code{m$meta} object. If uncertainty is taken into account,  the raw data used to obtain the estimates of TFR are stored as an additional element, called \code{raw_data.original}. A logical element \code{ar.phase2} indicates whether the autoregressive component of Phase II is considered in the estimation. In order to allow users to work with different subsets of countries with the same base of global estimates, information indicating whether the countries were processed separately has been also stored in the \code{meta} object. It is accessible via the \code{extra} element, created only if the \code{run.tfr.mcmc.extra} function has been invoked and if \code{uncertainty} is \code{TRUE}. Here, \code{extra_iter} and \code{extra_thin} are used to retrieve the settings for specific countries. The raw data in this case are stored in a list called \code{raw_data_extra}. It is overwritten every time \code{run.tfr.mcmc.extra} is called for the same country.

The results of Phase III are stored in the directory ``phaseIII''. It has the same structure as described above. It is generated either by the \code{run.tfr.mcmc} function in case of a one-step estimation, or by the  \code{run.tfr3.mcmc} function, in case of a two-step estimation. The meta file contains meta information related to the Phase III estimation. In the ``mc\textit{x}'' directories, the names of the hyperparameters and their notations for Phase III are listed in Table \ref{tbl1}. Similarly, the country-specific parameters and their notations are listed in Table \ref{tbl2}. All files in this case contain one value per (thinned) iteration. Note that the country-specific parameters for Phase III are only estimated for countries which are already in Phase III, which in our case is 41 countries.

\begin{table}[!htb]
	\centering
	\begin{tabular}{ccccc}
		\hline
		$\bar{\mu}$	& $\bar{\rho}$	 &$\sigma_\mu$	  & $\sigma_\rho$ & $\sigma_\varepsilon$ \\ \hline
		\code{mu}	& \code{rho} & \code{sigma.mu} & \code{sigma.rho} & \code{sigma.eps} \\ \hline
	\end{tabular}
	\caption{Country-independent parameters for Phase III in model \ref{modelphase3}, with their corresponding names in the code. They can be obtained using \code{tfr3.parameter.names()}.}\label{tbl1}
\end{table}

\begin{table}[!htb]
	\centering
	\begin{tabular}{cc}
		\hline
		$\mu_c$	& $\rho_c$	  \\ \hline
		\code{mu.c}	& \code{rho.c}  \\\hline
	\end{tabular}
	\caption{Country-specific parameters for Phase III in model \ref{modelphase3}, with their corresponding names in the code. They can be obtained using \code{tfr3.parameter.names.cs()}.}\label{tbl2}
\end{table}

The ``predictions'' directory is created by the \code{pop.predict} function and it holds binary files, one per country, each containing the predicted TFR trajectories for that country. 

Other convenience directories might have been created for speeding up processing. For example, the ``thinned\_mcmc\_9\_2100'' directory was created by \code{pop.predict} to hold the final chain for each parameter derived by applying the burnin, thinning and collapsing the three chains into one, in order to generate the predictions. Since we asked to generate 1,000 posterior TFR trajectories with burnin of 2,100 iterations, a thinning of 9 was applied to retrieve those trajectories: $3\cdot(5,100-2,100)/9 = 1,000$. Thus, the parameter files in the ``mc1'' subdirectory here all contain 1,000 rows. Note that these values will differ when working with a toy simulation. 

If functions for convergence diagnostics have been used, the simulation directory contains a folder ``diagnostics'' which holds results from these runs, one file per unique combination of thin and burnin.

%\section{Experiments, discussion, and future works}
\section{Experiments}
\label{experiments}
We have shown how the updated \pkg{bayesTFR} package can handle different versions of the TFR projection model.  In this section, we will present results of experiments under different settings and discuss the implications of these settings. Based on those experiments we will give recommendations for a reasonable configuration of the model. Finally, we will discuss future directions in the development of the package.

\subsection{Experiments with Settings}\label{sec:experiments}
The new version of \pkg{bayesTFR} allows the user to handle different types of modeling needs, summarized in Table~\ref{tbl0}. An analyst can choose between a five-year and an annual model, as well as between accounting for past uncertainty or not. Flexibility is added by allowing the user to treat vital registration (VR) records for selected countries as unbiased, as well as using the autoregressive component in Phase~II. 

However, a question of consistency of results between the various settings may arise. For example, a forecast should not change dramatically when switching from five-year to annual data. Currently, there are no annual observations collected for all countries, and only a few countries (such as New Zealand) have good annual vital registration data, the only available annual observations. Thus, if past uncertainty is not taken into account the model would be estimated on some version of interpolated data for most countries.

\subsubsection{Countries in Phase III with good records}
\begin{figure}[htb]
	\centering
	\begin{tabular}{cc}
		\includegraphics[width=0.5\textwidth]{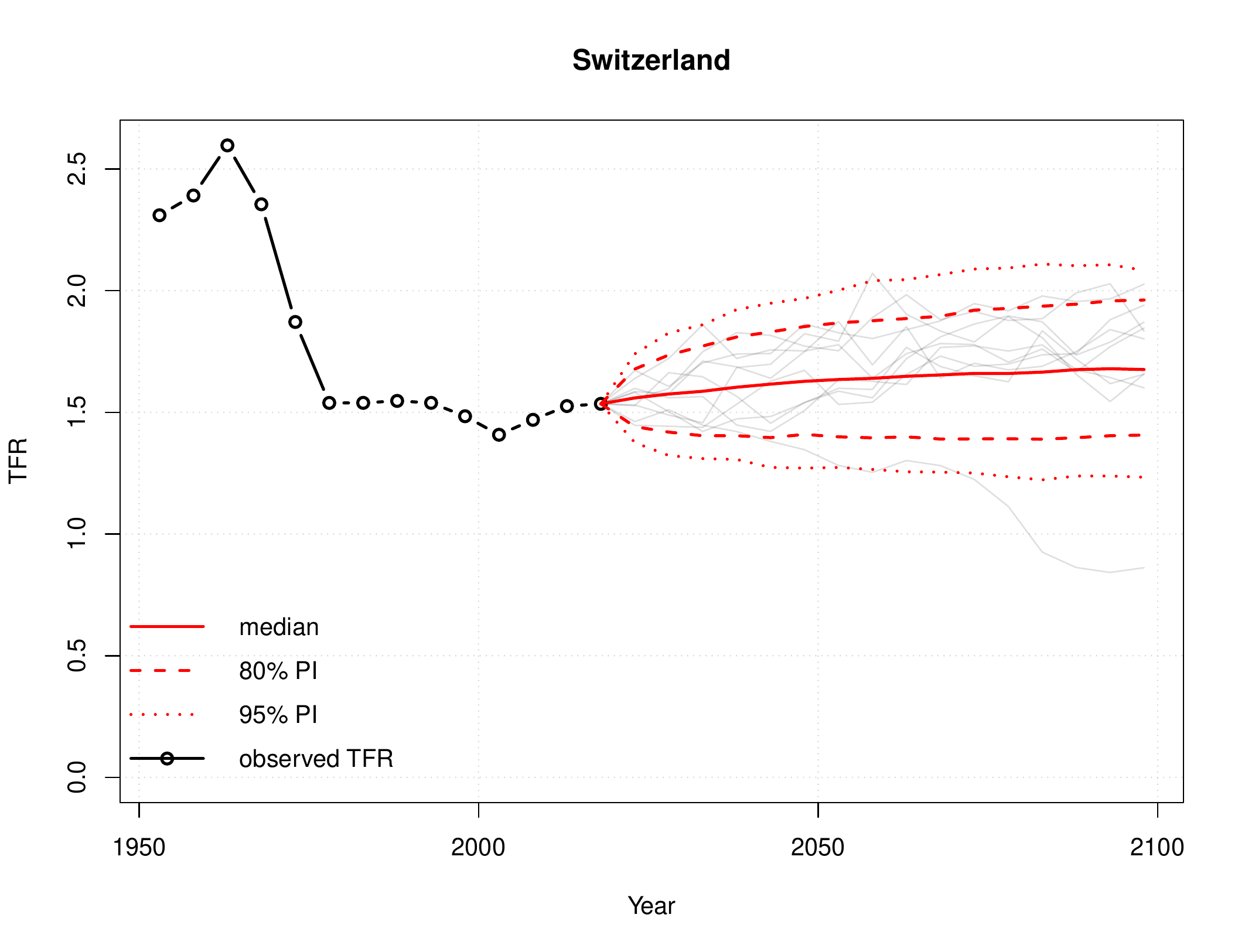} &
		\includegraphics[width=0.5\textwidth]{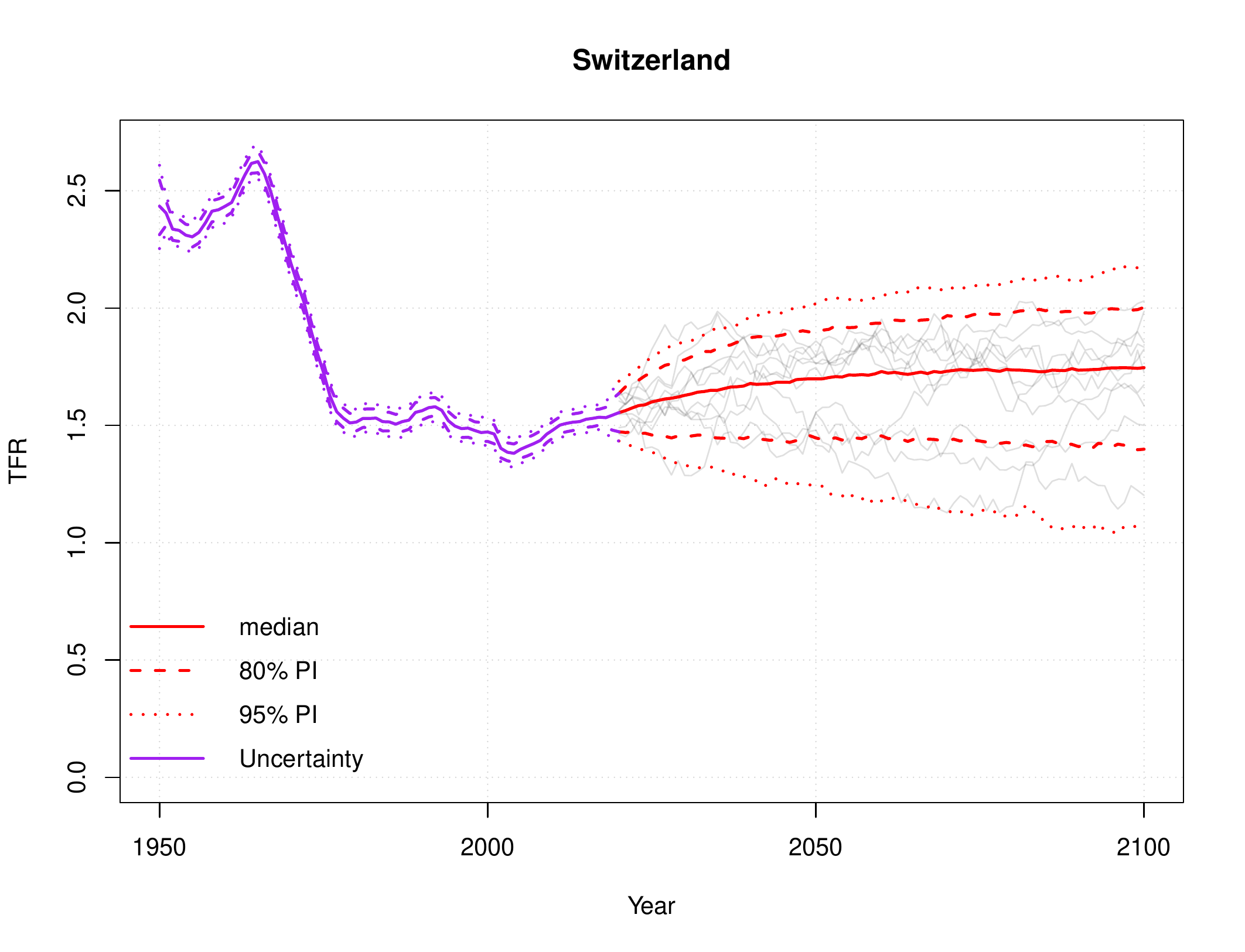}
	\end{tabular}
	\caption{TFR predictions for Switzerland. Left panel: 
		Original five-year model without accounting for past uncertainty. Right panel: Annual model with past uncertainty. }
	\label{fig-switzerland}
\end{figure}

The first major difference can be seen for countries in Phase III, especially for countries with high quality VR records. We take Switzerland as an example. The left panel of Figure~\ref{fig-switzerland} shows TFR projections for a five-year model without accounting for past uncertainty (cell D in Table~\ref{tbl0}), while the right panel shows results from an annual model with uncertainty about the past (cell A in Table~\ref{tbl0}). It can be seen that the results on the right yield wider probability intervals. For countries like Switzerland, the bias and uncertainty of past estimation is very low. Since the estimating process takes the linear interpolated TFR as the reference, the process can add extra bias to these data. Even though this is not large, the uncertainty propagated from the beginning of the forecast period could lead to a large difference.

\begin{figure}[htb]
	\centering
	\begin{tabular}{cc}
		\includegraphics[width=0.5\textwidth]{Figures/pred_p3_5} &
		\includegraphics[width=0.5\textwidth]{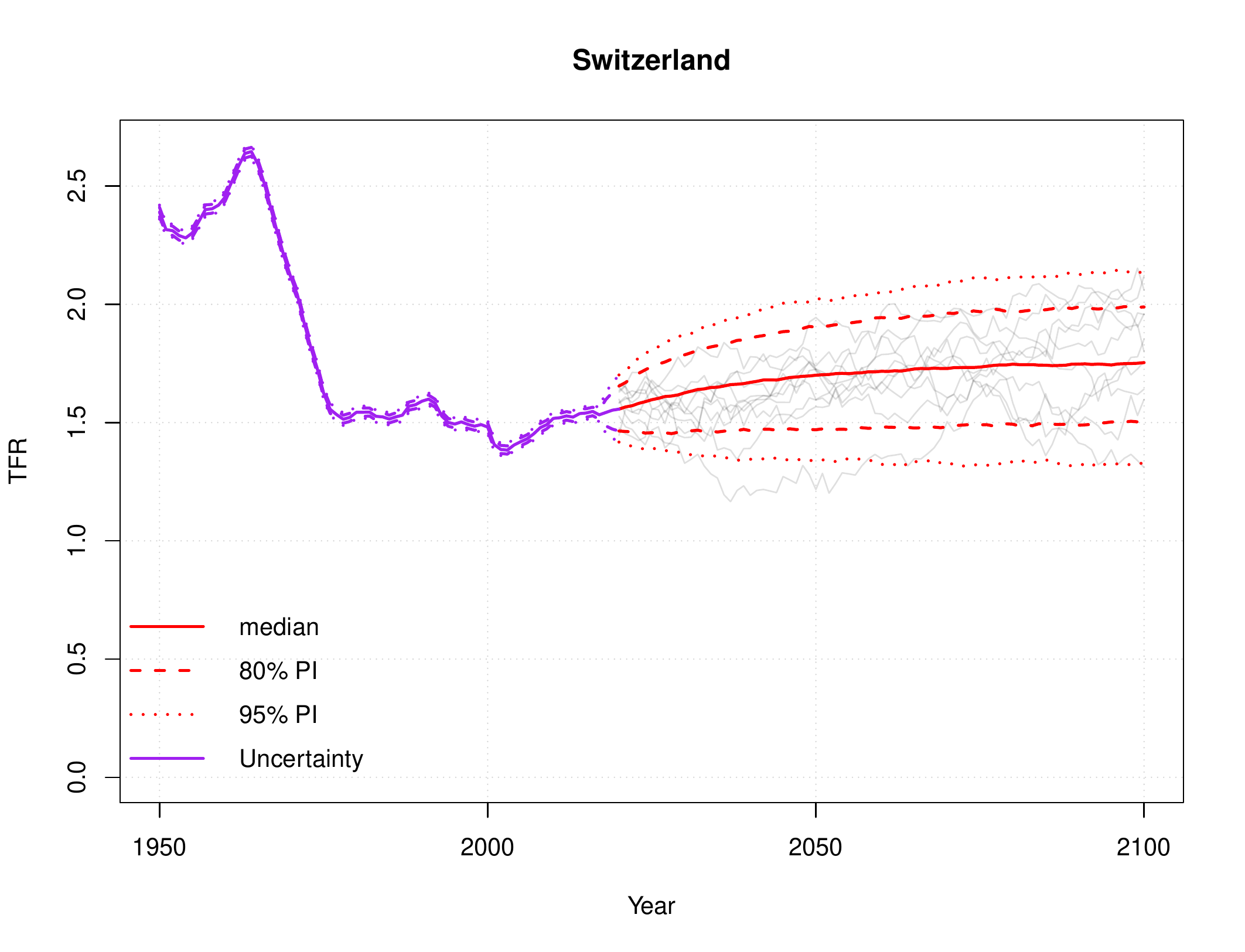}
	\end{tabular}
	\caption{TFR prediction of Switzerland. Left panel: 
		Original five-year model without accounting for past uncertainty. Right panel: Annual model with past uncertainty, with assuming VR records of selected countries (including Switzerland) as unbiased. }
	\label{fig-switzerland-2}
\end{figure}

Now we consider the VR records for a set of selected countries (OECD and some developed countries as unbiased; the list can be found in  the Appendix). The corresponding TFR projections for Switzerland are shown in the right panel of Figure~\ref{fig-switzerland-2}. 

It can be seen that when compared to results from a five-year model (left panel), the differences between the two sets of projections are negligible. It is important especially for countries  with nearly perfect historical data, such as Switzerland, that similar results be obtained whether annual or five-year data are used.

\subsubsection{Countries in Phase II}
\begin{figure}
	\centering
	\begin{tabular}{cc}
		\includegraphics[width=0.5\textwidth]{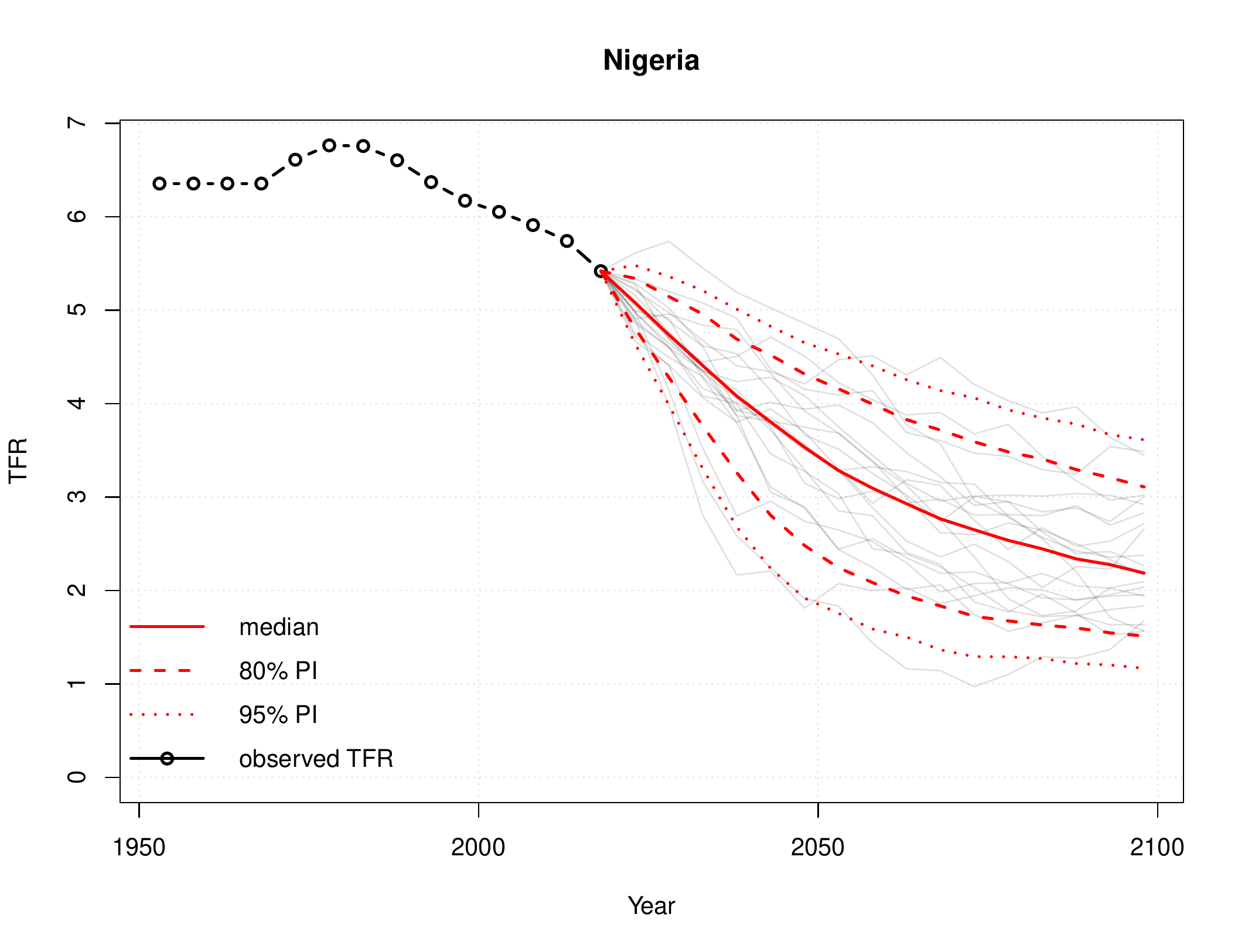} &
		\includegraphics[width=0.5\textwidth]{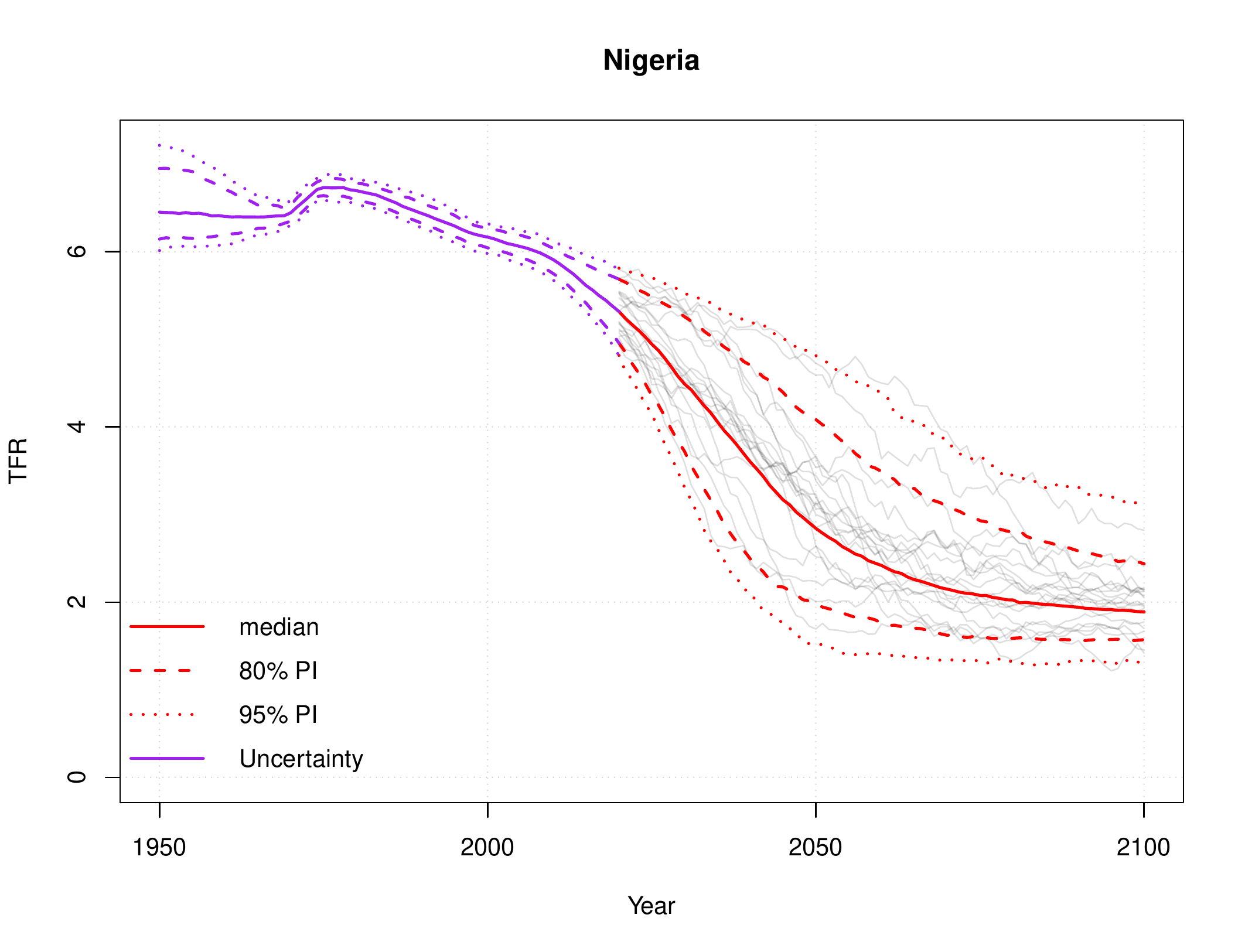}
	\end{tabular}
	\caption{TFR prediction of Nigeria. Left panel:
		Original five-year model without accounting for past uncertainty. Right panel: Annual model with past uncertainty without Phase II-AR(1). }
	\label{fig-nigeria}
\end{figure}

The second major difference relates to countries in Phase II, such as Nigeria. Figure \ref{fig-nigeria} shows the difference between a projection resulting from a five-year model without accounting for past uncertainty (left panel) and from an annual model with uncertainty about the past without applying the Phase II-AR(1) component. 

%However, for these countries, we may not argue that forecast with uncertainty should be similar with forecast without uncertainty. If we compare the predictions for Nigeria under two scenarios, Figure \ref{fig-nigeria} summarizes the difference.
\begin{figure}
	\centering\includegraphics[width=0.4\textwidth]{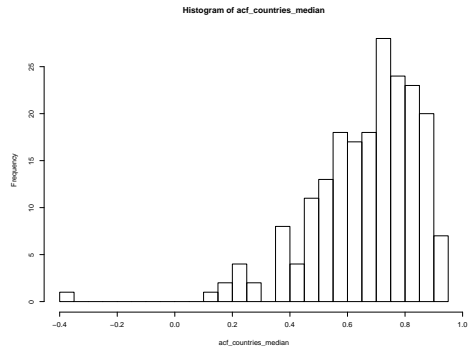}
	\caption{Histogram of autocorrelation for median Phase II residuals of all countries.}\label{acf}
\end{figure}

It can be seen that if we account for uncertainty and use annual data, the prediction shows a faster decline. Without performing an out-of-sample validation, it is impossible to say which of these projections is better. Nevertheless, a more detailed analysis revealed that the posterior median of the residuals $\varepsilon_{c,t}$ for all countries in model (\ref{model:old}) is highly autocorrelated. Figure \ref{acf} summarizes the estimates.

\begin{figure}
	\centering
	\begin{tabular}{cc}
		\includegraphics[width=0.5\textwidth]{Figures/pred_nigeria_annual} &
		\includegraphics[width=0.5\textwidth]{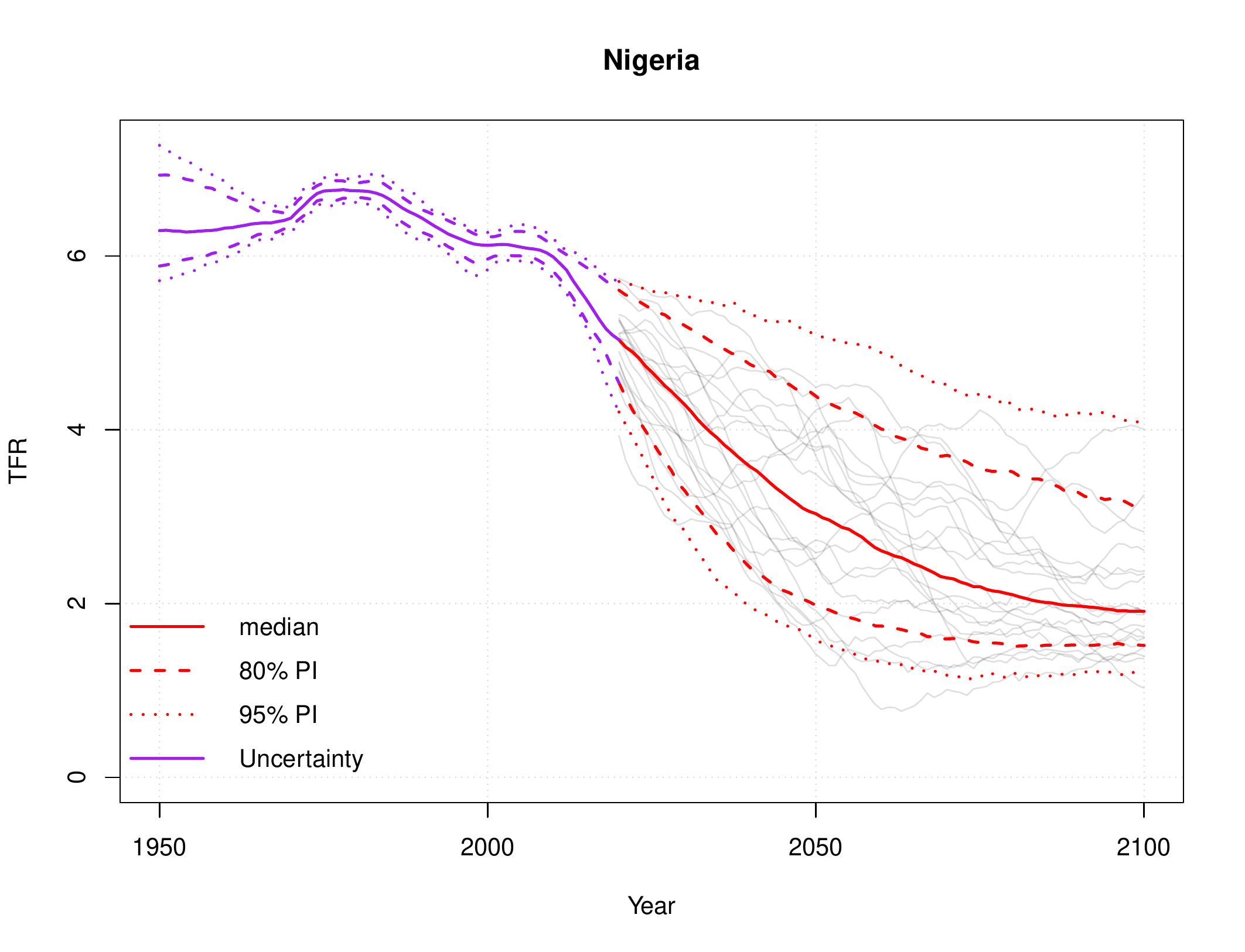}
	\end{tabular}
	\caption{TFR prediction for Nigeria resulting from an annual model with past uncertainty without Phase II-AR(1) (left panel) and with Phase II-AR(1) (right panel).}
	\label{fig-nigeria2}
\end{figure}

This suggests including the autocorrelation process in the modeling as defined in Equation~(\ref{model3}). Figure~\ref{fig-nigeria2} summarizes the differences.
The decline has become slower, which is more in line with the five-year projections.

\begin{figure}
	\centering
	\begin{tabular}{cc}
		\includegraphics[width=0.5\textwidth]{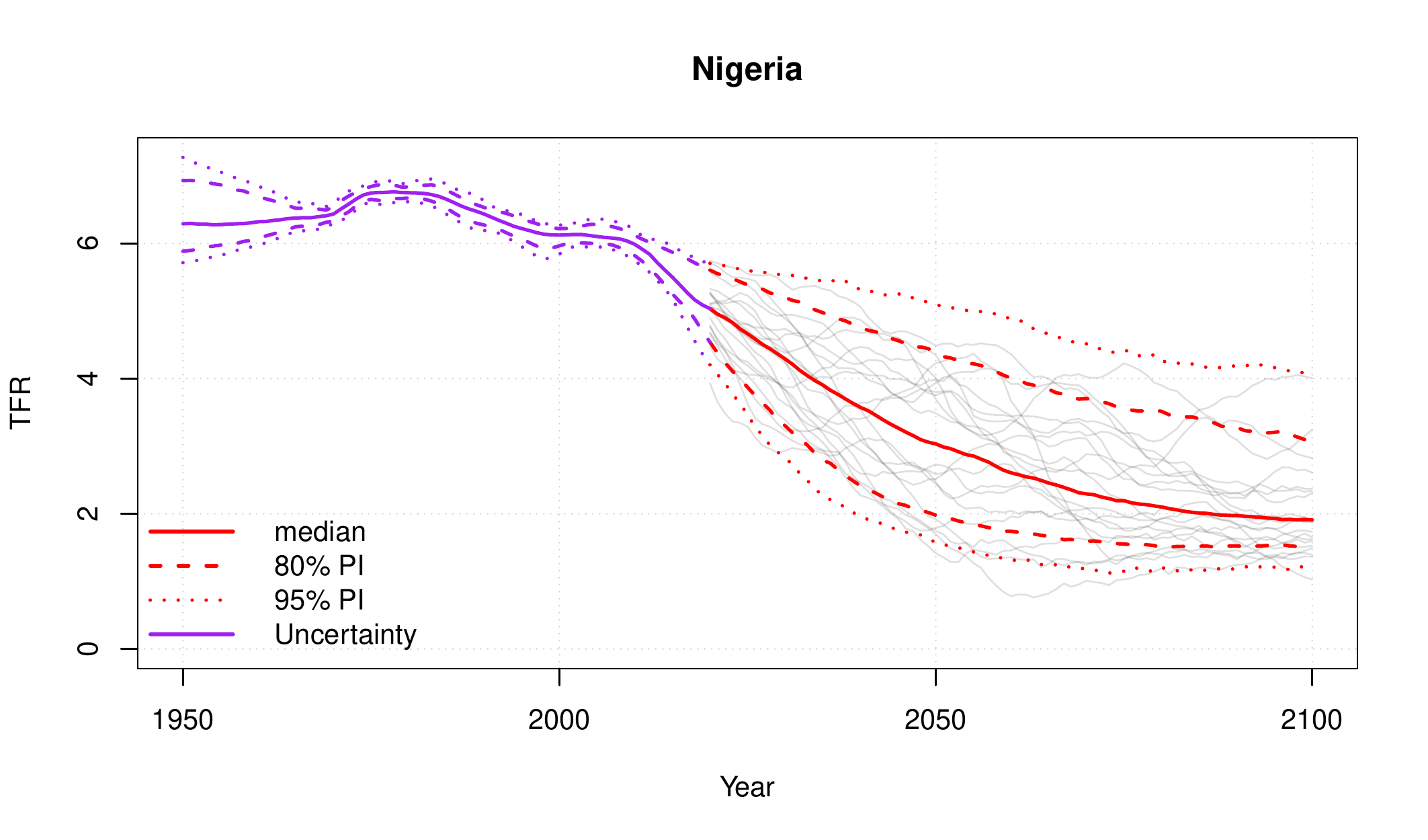} &
		\includegraphics[width=0.5\textwidth]{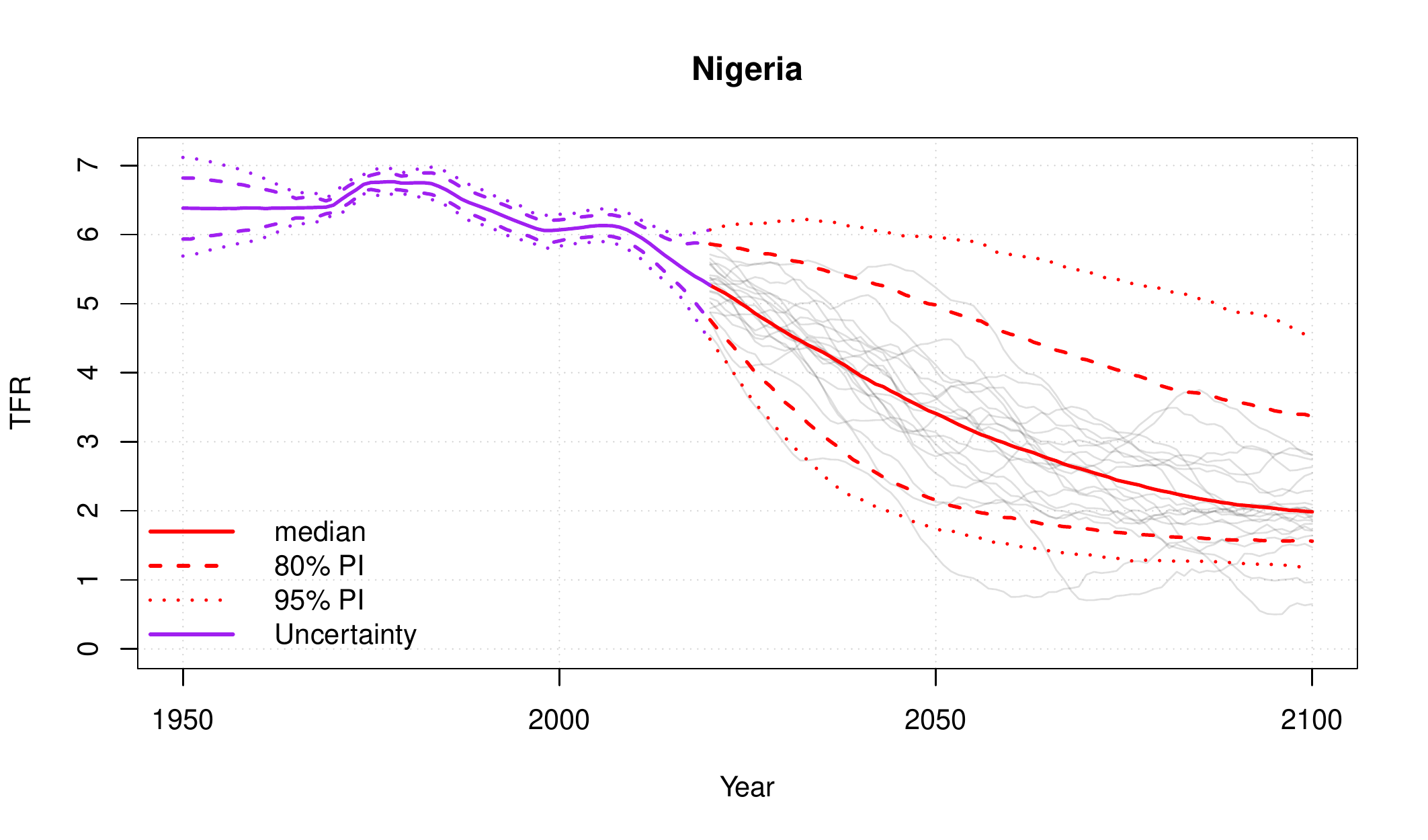} \\
		\includegraphics[width=0.5\textwidth]{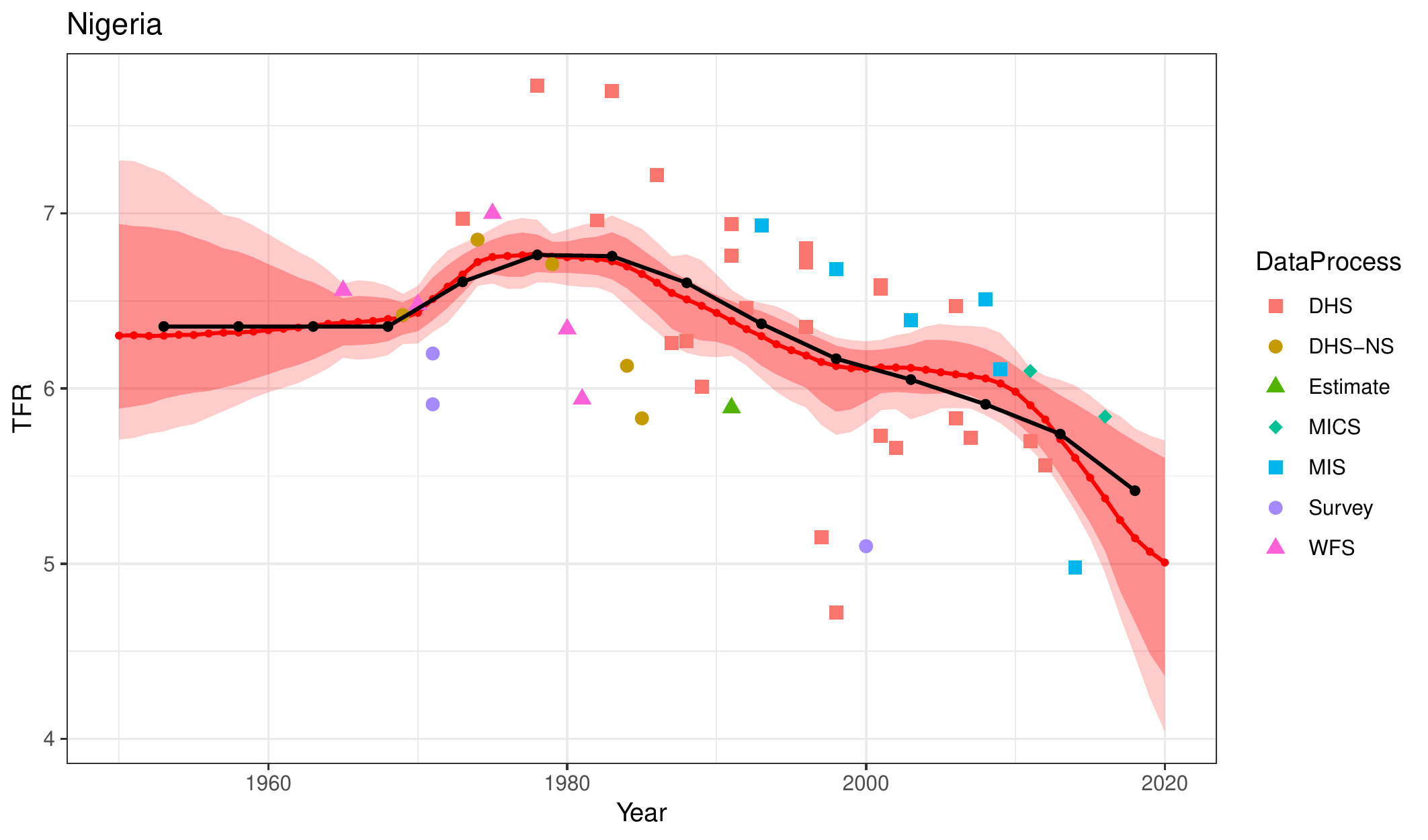} &
		\includegraphics[width=0.5\textwidth]{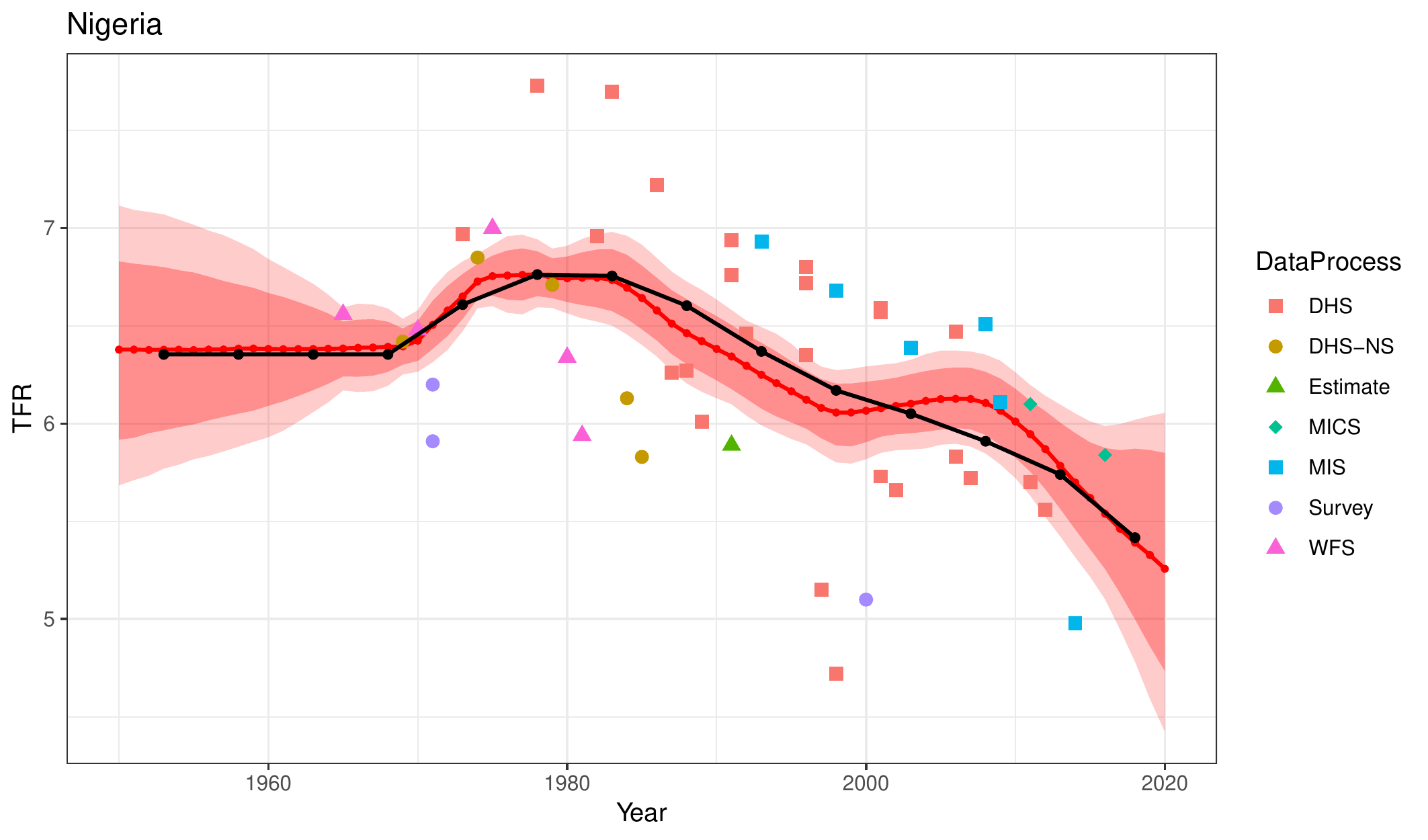}
	\end{tabular}
	\caption{TFR prediction (top row) and estimation (bottom row) for Nigeria from an annual model with uncertainty with autoregressive component. Left column: without lower bound on $\sigma_0$. Right column: with \code{sigma0.min = 0.04}. }
	\label{fig-nigeria3}
\end{figure}

It can be seen however, that the starting point of the projections (year 2020) is now lower, and in fact it is significantly lower than the current UN estimates. The standard deviation of $\varepsilon$ in model~(\ref{model3}) is less than 0.02, if the autoregressive component is included. This could be problematic, given that for developed countries with low TFR and relatively stable societies, the standard deviation of annual TFR changes is about twice as much as 0.02. This is likely a result of a possible smoothing of the data. To remedy that, we introduce a new lower bound on the $\sigma_0$ parameter (argument \code{sigma0.min} in \code{run.tfr.mcmc}) of 0.04, which becomes the new default. Figure \ref{fig-nigeria3} shows the relevant differences.

If the lower bound on $\sigma_0$ is applied, the prediction yields wider probability intervals as well as a higher median (top right panel), which better matches the five-year forecast. The estimation in this case (bottom right panel) also shows a better match with the raw data as well as with the UN estimates, which is another argument for using the new default for \code{sigma0.min}. 

\subsection{Recommendations}\label{sec:recommendation}
We have shown the flexibility of the new version of \pkg{bayesTFR} which can incorporate different variations of the TFR model as well as being compatible with the extant version of the model. As one of the key components in population projections currently adopted by the United Nations, this is a key step for migrating population projections from a five-year basis to an annual one. The package is designed to support UN analysts in this process, as well as to give other researchers and practitioners a tool to generate their own projections. 

In addition to incorporating past uncertainty of TFR in the forecast, and performing annual-based projections,  the package has introduced two other important components, namely the ability to specify vital registration data as unbiased, and the autoregressive component in Phase II. In Section~\ref{sec:experiments}, we have described the reasoning behind these two new options, as well as for setting a lower bound on the standard deviation. 

Based on our experiments and analysis, when using the annual model with uncertainty about the past in a production-like setting, i.e., if full convergence of the MCMC algorithm is desired, we recommend the following settings:
\begin{CodeInput}
R> annual <- TRUE
R> nr.chains <- 3
R> total.iter <- 62000
R> thin <- 10
R> burnin <- 2000
R> iso.goodvr <- c(36,40,56,124,203,208,246,250,276,300,352,372,
+    380,392,410,428,442,528,554,578,620,724,752,756,792,826,840)
R> m <- run.tfr.mcmc(output.dir = simu.dir.unc, nr.chains = 
+    nr.chains, iter = total.iter, annual = annual, thin = thin, 
+    uncertainty = TRUE, ar.phase2 = TRUE, 
+    iso.unbiased = iso.goodvr, parallel = TRUE)
R> pred <- tfr.predict(sim.dir = simu.dir.unc, end.year = 2100, 
+    burnin = burnin, nr.traj = 1000, uncertainty = TRUE)
\end{CodeInput}

The ISO codes listed include most European countries, Australia, Japan, South Korea, New Zealand and the United States. These countries have a long history of vital registration with coverage rates often around 99\%, indicating that their observations have been of high quality.

You should expect a full simulation with these settings to run for several days. Thus, we recommend processing it by a batch script in the background, so that it can be left unattended.

\section{Discussion}\label{discussion}
In this article, we have described the latest major update of the \proglang{R} package \pkg{bayesTFR}. This update significantly enriches the modeling framework in the previous version of the package, and gives analysts the flexibility to account for past TFR uncertainty, use annual data, and allow for an autoregressive model in Phase~II. Moreover, by making use of the vectorization nature of \proglang{R}, the computational cost has been kept at a reasonable level while making the model more sophisticated. New functions for visualizing estimation results, as well as updated analysis tools will further support analysts in exploring the package outputs.

On the package development side, there are at least two major areas for future improvements.
The first is modeling age-specific fertility rates with past uncertainty which is of interest to demographers. The second would be further vectorizing the MCMC process. If past uncertainty is included in the model, updating the estimates of TFR is the most time-consuming part of the process. Since we consider each past TFR per country and time period as a parameter, it  adds over 14,000 parameters in the annual case. Thus, the speed of the Metropolis-Hastings step for updating TFR plays a big role in determining the overall speed of the method. If past uncertainty is not included, updates of country-specific parameters dominate the computing time, and thus are subject to further optimization.

On the modeling side, there are also two obvious directions for improvement. 
First, instead of modeling the bias and standard deviation based on linear regression for each country separately, these could be folded into the process, giving a fully united probabilistic model.  
A pooled version could yield more robust estimates, especially given the 
small amount of data in some surveys.
Another direction is related to the completeness of the VR data. The completeness of VR coverage is the most important factor for how precise the VR records are, and this is an important consideration for VR but not for other surveys. Due to the low bias of high quality vital registration systems, more research could be done on how to incorporate this information in the model.

\section*{Acknowledgments}
This research was supported by NICHD grant no. R01 HD070936. Its content is solely the responsibility of the authors and does not necessarily represent the official views of NICHD or the United Nations. The authors are grateful to Patrick Gerland and Nathan Welch for helpful discussions and insightful comments. 

\bibliography{jss4235}
\clearpage

\appendix
\section*{Prior Distributions}
Here we provide a full description of the Bayesian hierarchical model,
which was summarized in the main text for annual model. Level 1 is used if \code{uncertainty = TRUE}:
\begin{align*}\label{eq:model}
\text{Level 1: }
& y_{c,t,s} | f_{c,t} \sim \mathcal{N}(f_{c,t} + \delta_{c,s}, \rho_{c,s}^2)\, , \\
& \mathbb{E}[\delta_{c,s}] = \bm{x}_{c,s}\bm{\beta} \,,\\
& \mathbb{E}[\rho_{c,s}] = \bm{x}_{c,s}\bm{\gamma}\,;\\
\text{Level 2: }
&\text{Phase I: }f_{c,t} = f_{c,t-1} + \varepsilon_{c,t} \,, \\
&\text{Phase II: }f_{c,t} = f_{c,t-1} - d_{c,t-1} \,, \\
&\texttt{ar.phase2=FALSE: } d_{c,t} = g_{c,t} + \varepsilon_{c,t}\,,\\
&\texttt{ar.phase2=TRUE: }d_{c,t} - g_{c,t} = \phi(d_{c,t-1}-g_{c,t-1}) + \varepsilon_{c,t} \,,\\
&\text{Phase III: }f_{c,t} = \mu_c + \rho_c(f_{c,t-1} - \mu_c) + \varepsilon_{c,t} \,,\\
&\bm{\theta}_c = (\Delta_{c1}, \Delta_{c2}, \Delta_{c3}, \Delta_{c4}, d_c)\,\\
& \varepsilon_{c,t} \sim \mathcal{N}(0, \sigma_{c,t}^2) \,,\\
& \begin{aligned}g(f_{c,t}|\bm{\theta}_c) = &-\frac{d_c}{1 + \exp\left(-\frac{2\ln(9)}{\Delta_{c1}}\left(f_{c,t} - \sum_i\Delta_{ci} + 0.5\Delta_{c1} \right)\right)} \\ 
& + \frac{d_c}{1 + \exp\left(-\frac{2\ln(9)}{\Delta_{c3}}\left(f_{c,t} - \Delta_{c4} - 0.5\Delta_{c3} \right)\right)}\end{aligned}
\end{align*} 

The country-specific variance, $\sigma_{c,t}$, varies according
to the phase and the current fertility level, as follows:
\begin{align*}
& \sigma_{c,t} = c_{1975}(t)\left(\sigma_0 + (f_{c,t} - S)(-aI_{f_{c,t} > S} + b I_{f_{c,t} < S}) \right) \text{  for }t\text{ is in Phase II.}\\
& c_{1975}(t) = c I_{t\leq 1975} + I_{t>1975}\,.
\end{align*}

The country-level parameters, $\{U_c, \rho_c, \mu_c, \gamma_{ci}, \Delta_{c4}, d_c\}$,  are specified as follows:
\begin{align*}
\text{Level 3: }
& U_c \left\{\begin{aligned}
& =f_{c,\tau} &&\quad \tau_c\geq 1950\\
& \sim U(\min\{5.5, \max_t\{f_{c,t}\}\}, 8.8) &&\quad \tau_c < 1950
\end{aligned} \right. \\
& \phi_c = \log\left(\frac{d_c - 0.05}{0.5 - d_c} \right)\,,\\
& \phi_c \sim \mathcal{N}(\chi, \psi^2)\,,\\
& \Delta_{c4}' = \log\left(\frac{\Delta_{c4} - 1}{2.5 - \Delta_{c4}} \right)\,,\\
& \Delta_{c4}' \sim \mathcal{N}(\Delta_4, \delta_4^2)\,,\\
& p_{ci} = \frac{\Delta_{ci}}{U_c - \Delta_{c4}}\text{ for }i = 1,2,3 \,,\\
& p_{ci} = \frac{\exp(\gamma_{ci})}{\sum_j\exp(\gamma_{cj})} \,,\\
& \gamma_{ci} \sim \mathcal{N}(\alpha_i, \delta_i^2)\,,\\
& \mu_c \sim  \mathcal{N}(\bar{\mu}, \sigma_\mu^2)\,, \\
& \rho_c \sim \mathcal{N}(\bar{\rho}, \sigma_\rho^2)\,;
\end{align*} 
where $\tau_c$ is the starting year of phase II for country $c$.
%which is determined priorly. 

The hyperparameters are $\{s_\tau, \sigma_0, a, b, S, c, \sigma_\epsilon, \chi, \psi, \Delta_4, \delta_4, \bm{\alpha}, \bm{\delta}, \bar\mu, \sigma_\mu, \bar{\rho}, \sigma_\rho \}$. Some of these refer to Level 2 and some to Level 3. 
The prior distribution of these hyperparameters is as follows ($\phi$ is used if \code{ar.phase2 = TRUE}):
\allowdisplaybreaks
\begin{align*}
\text{Level 4: }
1/s_\tau^2 &\sim \text{Gamma}(1, 0.4^2) \,,\\
\sigma_0 &\sim  U[0.002, 0.6], \quad \text{recommended } U[0.04, 0.6] \,, \\
a &\sim U[0, 0.2]\,,\\
b &\sim U[0, 0.2]\,,\\
S &\sim U[3.5, 6.5]\,,\\
c &\sim U[0.8, 2]\,,\\
\sigma_\epsilon& \sim U[0, 0.5] \,,\\
\chi &\sim  \mathcal{N}(-1.5, 0.6^2) \,,\\
1/\psi^2 &\sim \text{Gamma}(1, 0.6^2)\,, \\
\Delta_4 & \sim \mathcal{N}(0.3, 1)\,, \\
1/\delta_i^2 & \sim \text{Gamma}(1, 1) \text{ for }i = 1,2,3,4\,, \\
\alpha_1 & \sim \mathcal{N}(-1, 1) \,,\\
\alpha_2 & \sim \mathcal{N}(0.5, 1) \,,\\
\alpha_3 & \sim \mathcal{N}(1.5, 1) \,,\\
\bar{\mu} & \sim U[0, 2.1]\,,\\
\sigma_\mu & \sim U[0, 0.318]\,,\\
\bar{\rho} & \sim U[0, 1]\,,\\
\sigma_\rho & \sim U[0, 0.289]\,,\\
\phi & \sim U[0,1]\,.
\end{align*}

\section*{List of Unbiased VR countries}
In this article, we are assuming that the following countries have nearly perfect VR histories:

Australia, Austria, Belgium, Canada, Czech Republic, Denmark, Finland, France, Germany, Greece, Iceland, Ireland, Italy, Japan, Korea, Latvia, Luxembourg, Netherlands, New Zealand, Norway, Portugal, Spain, Sweden, Switzerland, Turkey, the United Kingdom, the United States.
\end{document}